\documentclass[acmlarge,authorversion]{acmart}

\usepackage[nolist,nohyperlinks]{acronym}
\usepackage[thinspace,thinqspace,squaren]{SIunits} %siunitx :'(
\usepackage{soul}
\usepackage{subfigure}
\usepackage{hyperref}

\AtBeginDocument{%
  \providecommand\BibTeX{{%
    \normalfont B\kern-0.5em{\scshape i\kern-0.25em b}\kern-0.8em\TeX}}}

\setcopyright{acmlicensed}
\acmJournal{IMWUT}
\acmYear{2021}
\acmVolume{5}
\acmNumber{4}
\acmArticle{157}
\acmMonth{12}
\acmPrice{15.00}
\acmDOI{10.1145/3494986}

\acmSubmissionID{8993}

\begin{acronym}
\acro{AR}[AR]{augmented reality}
\acro{HAR}[HAR]{human activity recognition}
\acro{EMG}[EMG]{electromyographic}
\acro{IMU}[IMU]{inertial measurement unit}
\acro{PIN}[PIN]{personal identification number}
\acro{PLI}[PLI]{power-line noise}
\acro{IIR}[IIR]{infinite impulse response}
\acro{CNN}[CNN]{convolutional neural network}
\acro{CRNN}[CRNN]{convolutional recurrent neural network}
\acro{LSTM}[LSTM]{long short-term memory}
\acro{TSC}[TSC]{time series classification}
\acro{XR}[XR]{extended reality}
\acro{CWT}[CWT]{continuous wavelet transformation}
\acro{ReLU}[ReLU]{rectified linear unit}
\acro{DTW}[DTW]{dynamic time warping}
\end{acronym}

\begin{document}

\title{My(o) Armband Leaks Passwords: An EMG and IMU Based Keylogging Side-Channel Attack}

\author{Matthias Gazzari}
\authornote{Both authors contributed equally to this research.}
\orcid{0000-0002-8656-8320}
\affiliation{%
  \institution{Technical University of Darmstadt}
  \department{Secure Mobile Networking Lab}
  \city{Darmstadt}
  \country{Germany}
}
\email{mgazzari@seemoo.tu-darmstadt.de}
\author{Annemarie Mattmann}
\authornotemark[1]
\orcid{0000-0003-0118-0747}
\affiliation{%
  \institution{Technical University of Darmstadt}
  \department{Secure Mobile Networking Lab}
  \city{Darmstadt}
  \country{Germany}
}
\email{amattmann@seemoo.tu-darmstadt.de}
\author{Max Maass}
\orcid{0000-0001-9346-8486}
\affiliation{%
  \institution{Technical University of Darmstadt}
  \department{Secure Mobile Networking Lab}
  \city{Darmstadt}
  \country{Germany}
}
\email{mmaass@seemoo.tu-darmstadt.de}
\author{Matthias Hollick}
\orcid{0000-0002-9163-5989}
\email{mhollick@seemoo.tu-darmstadt.de}
\affiliation{%
  \institution{Technical University of Darmstadt}
  \department{Secure Mobile Networking Lab}
  \city{Darmstadt}
  \country{Germany}
}

\renewcommand{\shortauthors}{Gazzari and Mattmann, et al.}

\begin{abstract}
Wearables that constantly collect various sensor data of their users increase the chances for inferences of unintentional and sensitive information such as passwords typed on a physical keyboard.
We take a thorough look at the potential of using \ac{EMG} data, a sensor modality which is new to the market but has lately gained attention in the context of wearables for \ac{AR}, for a keylogging side-channel attack. Our approach is based on neural networks for a between-subject attack in a realistic scenario using the Myo Armband to collect the sensor data. In our approach, the \ac{EMG} data has proven to be the most prominent source of information compared to the accelerometer and gyroscope, increasing the keystroke detection performance.
For our end-to-end approach on raw data, we report a mean balanced accuracy of about \unit{76}{\%} for the keystroke detection and a mean top-3 key accuracy of about \unit{32}{\%} on 52 classes for the key identification on passwords of varying strengths.
We have created an extensive dataset including more than 310\,000 keystrokes recorded from 37 volunteers, which is available as open access along with the source code used to create the given results.
\end{abstract}



\maketitle

\section{Introduction}

Wearables are designed to gather and process information for the user's benefit.
They are also designed to be worn for most if not all of the day.
Thus, they will inevitably be present when the wearers engage in sensitive activities, gathering data that may contain traces of these activities.
% previous attacks based on smart watches (all physical keyboard attacks are between-subject)
Human-centric sensors embedded in ubiquitous wearable devices such as smartwatches have been shown to pose risks for keylogging side-channel attacks to snoop on human input on physical keyboards \cite{wang2015mole, liu2015good, maiti2016smartwatch, pandelea2019password}, keypads \cite{liu2015good, wang2016friend, liu2018aleak}, or virtual keyboards on touch screens \cite{sarkisyan2015wristsnoop, maiti2015smart, maiti2018side}.
% dictionary based attacks vs attacking unstructured text
Successful attacks based on this data can leak confidential information such as text typed on a keyboard by matching it against English words from dictionaries \cite{wang2015mole, liu2015good, maiti2016smartwatch}.
Preliminary work indicates that such approaches could also be used to infer unstructured text such as passwords \cite{pandelea2019password}.

% clarify why EMG and EMG+IMU is important/relevant compared to prior work focusing on IMU
Most prior work focuses on \ac{IMU} sensor data, particularly on the accelerometer and the gyroscope.
However, over the past years, more and more wearables have entered the consumer market equipped with a multitude of different sensors like \ac{EMG} sensors.
Devices that record \ac{EMG} data can capture the muscle activity responsible for moving individual fingers and thus offer a novel attack vector by providing opportunities for a more fine-grained detection and differentiation of typing different keys on the keyboard.
Typically, those wearables also include an \ac{IMU}, providing the opportunity to complement attacks based on \ac{EMG} with information about the coarse movements of the arm \cite{Zhang2017}.
Given these considerations and the results of prior research, we suggest an approach utilizing both \ac{IMU} and \ac{EMG} sensor data both in isolation and combination to evaluate the importance of the \ac{EMG} data, which may imply a greater security risk of using wearables equipped with both sensors compared with those that only offer an \ac{IMU} sensor.

% provide EMG based examples/state of the art
Facebook recently proposed a personalized virtual keyboard based on a custom \ac{EMG} wrist band for \ac{XR} applications \cite{website:facebook}, using the sensor data to interpret typing of fingers on an imaginary keyboard.
Their work is similar to prior research of using the \ac{EMG} sensors of the Myo armband for a virtual T9 keyboard by inferring nine different finger gestures \cite{Fu2020}, as well as to prior work of using a custom \ac{EMG} sensor exploring text entry with 16 keys of the left hand on a keyboard printed on a paper and a physical keyboard.
These kinds of approaches open up new possibilities for user interaction, but also highlight the possible impacts of abusing \ac{EMG} as well as \ac{IMU} data as a potential side-channel for snooping on user input.

% possible attack scenarios: targets beyond the device connected to the wearable, based on app or cloud data
Compared to a keylogger installed on a single computing device by a malicious actor, inferring keystrokes from sensor data collected by wearables opens up opportunities to attack not only a single computer but as many as a victim is interacting with while wearing these devices.
Like smartwatches, such wearables may be connected to mobile devices or operate independently, for example, in an \ac{AR} context.
% convenience (especially relevant for EMG devices as these have to be 'synchronized')
For reasons of convenience or simply unawareness of potential consequences, these wearables may not be taken off when engaging in sensitive activities such as typing on a physical keyboard.
Thus, sensor data of such wearables may reveal everything that has been typed on a personal, work or any other device.
This is especially the case for wearables based on \ac{EMG} sensors, as those typically involve a more time-consuming setup.
Furthermore, this poses an issue in scenarios in which users depend on their wearables, as is the case for prostheses based on \ac{EMG} \cite{visconti2018} and medical monitoring \cite{majumder2019smartphone,dunn2018wearables}.

In order to gain access to the sensor data, an attacker could use a malicious app installed on a single device connected to the wearable, enabling the adversary to infer keystrokes from every device the victim is typing on.
This app could either be installed by the adversary or by the victims themselves, potentially disguised as a harmless application for using the wearable in a legitimate use-case.
Alternatively, with sensor data stored in the cloud, an adversary could access the sensor data of multiple victims directly, without having to interact with the victims.
Such cloud storage may be legitimately used by a service providers of the wearable to improve their service, but offers a larger attack surface, possibly exposing the sensor data of all users of their service.
% user not aware of danger: 1. doesn't know what EMG is, 2. device capabilities unknown because advertised for something different or coarse input
Furthermore, users may also willingly share their data with third parties, unaware what could be inferred from sensors like \ac{EMG}.
They may underestimate the risk due to these devices being advertised for more coarse types of input like controlling certain applications with gestures.

To the best of our knowledge, we are among the first to look into keylogging side-channel attacks based on forearm \ac{EMG} and \ac{IMU} sensor data for inferring unstructured text like passwords.
One of the \ac{EMG} sensors that have been sold on the consumer market is the \emph{Myo armband}, a wearable device worn on the lower arm and connected to a computer via Bluetooth. It records electromyographic data, i.e.\ muscle data, and \ac{IMU}, i.e.\ accelerometer, gyroscope and magnetometer data, to recognize gestures and finger movement.
Prior to us, \citet{Zhang2017} have used the \ac{EMG} data of the Myo armband for keystroke detection and finger differentiation on individual users (\emph{within-subject}).
However, their approach is limited to touch typists in a controlled setting and does not support generalization between users.

% our work/contributions
In our work, we use the raw \ac{EMG} and \ac{IMU} (accelerometer and gyroscope) data from two Myo armbands worn on both lower forearms to infer keystrokes typed on a physical keyboard in a \emph{between-subject} scenario to infer keystrokes from previously unknown users.
We assume an adversary who has access to either the \ac{EMG}, accelerometer or gyroscope data (or a combination thereof) of the victim, for example, through the use of a malicious application installed on the device to which the Myo armband is connected.
The adversary trains a neural network in advance with a custom dataset of labeled sensor data, based on the input of typists other than the victim.
In contrast to within-subject approaches, the adversary does not need to create an individual training set by observing the victim prior to executing the attack.
Instead, the adversary can use the trained model to infer the typing of the victim based only on the gathered sensor data, leveraging the generalization capabilities of the neural network approach.
In doing so, the adversary seeks to gain information on what has been typed on a physical keyboard by the victim, including text and passwords.
By hiring as many different typists as possible for the original dataset, the adversary can further improve the performance.

We compare the influence of each individual sensor with the performance of a combination of all sensors to evaluate the potential security impact of adding \ac{EMG} sensors to wearable devices.
We try to mimic the natural environment of the typists, including its changes and the typists' natural behavior, by setting minimal constraints on the typists to gain representative results and evaluate a realistic scenario for a keylogging side-channel attack. Furthermore, we include touch and hybrid typists in our analysis. In order to investigate the influence of different passwords, we include test data for passwords of different strengths in our evaluation, ranging from popular but insecure to randomly generated passwords and long passwords based on multiple words.
To infer the keystrokes, we apply and compare four different supervised end-to-end machine learning models for (1) detecting the presence of keystrokes and (2) identifying the key which has been typed. One of these four models is the state of the art end-to-end neural network model for time series classification taken from \citet{fawaz2019}.
As part of this work, we publish the first multimodal \ac{EMG} and \ac{IMU} dataset, including the timestamped keystroke events, as well as the source code for training the machine learning models along with the respective analyses to support further research in this area.

Thus, our main contributions are the following:
\begin{itemize}
	\item We collect and release the first large open access dataset composed of 310\,566 keystrokes (about \unit{34}{\hour} of typing) in total with the corresponding \ac{EMG} and \ac{IMU} data of two Myo armbands, recorded from 37 participants in a realistic scenario.
	\item We are the first to implement a between-subject approach to generalize between users for keystroke inference of words as well as unstructured text like passwords, based on \ac{EMG} sensor data, and to evaluate the importance of \ac{EMG} compared to \ac{IMU} data in this context.
	\item We evaluate four different, end-to-end trained neural network architectures for keystroke detection and identification in order to compare their performance and avoid a bias introduced by a certain model structure. This is the first application of end-to-end neural networks to implement a human-centric sensor-based keylogging side-channel attack.
\end{itemize}

% road map
This paper is structured as follows: In Section~\ref{sec:datastudy}, we outline the study design and describe the resulting dataset.
Section~\ref{sec:implementation} provides details on our classification approach, including the involved data pre-processing, the model descriptions, the training process and the post-processing. We also include the results of the keystroke detection and key identification classifiers.
Section~\ref{sec:discussionlimits} includes a short discussion of the results and lists the limitations of our approach.
In Section~\ref{sec:relatedwork}, we give a comparison to related work.
Section~\ref{sec:conclusion} provides a short conclusion for our work.

%%%%%%%%%%%%%%%%%%%%%%%%%%%%%%%%%%%%%%%%%%% Data Study %%%%%%%%%%%%%%%%%%%%%%%%%%%%%%%%%%%%%%%%%%%%%

\section{Data Study}\label{sec:datastudy}

Inferring keystrokes from the data recorded by wearable devices involves two steps: \emph{Keystroke detection} and \emph{key identification} \cite{monacosok}. Keystroke detection refers to detecting the presence and timing of any keystroke by determining the time of either a press event or a release event, or by inferring the key state at any given time. Key identification refers to determining which physical key was pressed given a detected keystroke.
The keystroke detection could be preceded by a typing activity recognition, i.e.\ detecting those sequences within the constant stream of data recorded by a wearable device that actually contain keystrokes. However, this would require recording multiple daily activities along with typing on a keyboard, which is out of scope for this project.

In this work, we focus on keystroke detection and key identification. Thus, for recording the data we need in order to train and test our classifiers, we assume a typing scenario in which the participants either sit at or stand in front of a table, ready to write on the keyboard. At the same time, we do not prohibit movements by the participants that would usually occur while typing on a keyboard, such as drinking or moving the body. In doing so, we require our keystroke detection models to distinguish between typing and non-typing movements within a typing scenario.

\subsection{Data}

The data we record includes the \emph{sensor data} recorded by the two Myo armbands, the \emph{ground truth} recorded by the laptop while typing on the external physical keyboard and the \emph{meta data} associated with each recording.
The dataset is available as open access at \url{https://doi.org/10.5281/zenodo.5594651} and the source code used for recording this dataset is available as free software at \url{https://github.com/seemoo-lab/myo-keylogging}.

\begin{figure}
\centering
\includegraphics[width=0.5\textwidth]{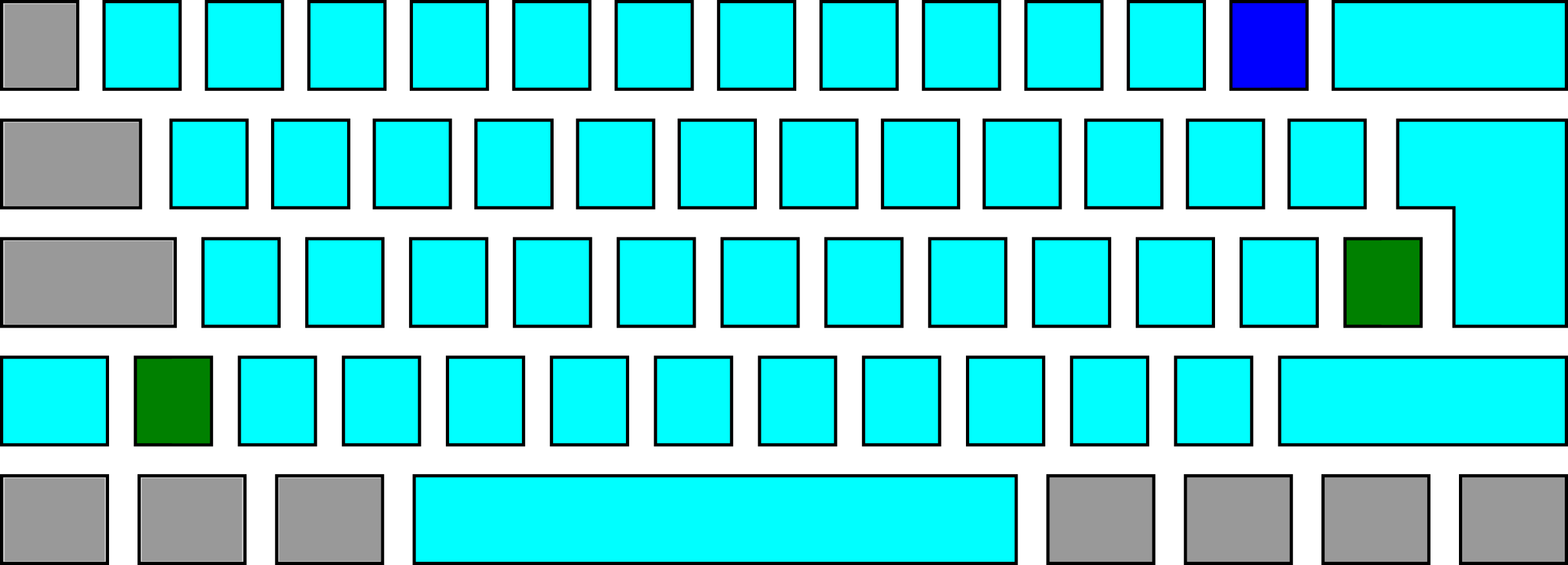}
\Description{A sketch showing the blank keys of a \unit{60}{\%} ISO keyboard, colored to indicate the considered keys in this study. Grey keys are all modifiers except for left and right shift, the blue key is the one left to backspace and the green keys are the ones on the right to left shift and on the lower left to the \texttt{return} key.}
\caption{The 52 keys of an ISO keyboard considered for the classification. Grey keys are not considered, green keys are only included in recordings with a German layout and the blue key is only used in recordings with a US layout.}
\label{fig:keyboard_keys}
\end{figure}

The sensor data includes both the raw \ac{EMG} and \ac{IMU} (accelerometer, gyroscope and magnetometer) data, recorded at \unit{200}{\hertz} and \unit{50}{\hertz} respectively. As the Myo armband does not attach timestamps to the data, these are added as soon as the Bluetooth packets arrive on the laptop. The ground truth consists of timestamped keystrokes recorded by the laptop while typing on the physical keyboard. For our work we consider a total of 52 keys of an ISO keyboard, with either a German (51 keys) or a US (50 keys) logical keyboard layout as shown in Fig.~\ref{fig:keyboard_keys}. Both, the sensor and the keystroke data, are used to train and evaluate the classifiers in a supervised learning fashion. The keystrokes refer to physical keys (and not the assigned characters) to ensure that our approach is invariant against logical keyboard layout changes. The meta data includes information such as a participant's unique identifier and the type of the recording.

One \emph{recording session} is comprised of multiple \emph{recordings}, each of which is associated with a certain task. This allows for a simple rerun of a task if a problem occurs during the recording and for easily switching between task types to make the recording more varied for the participants. One participant may or may not take part in multiple recording sessions. All recording sessions are represented by unique identifiers in the meta data and all recordings have unique identifiers in regard to the respective recording session.

\subsection{Experimental Setup}

\begin{figure}
\centering
\includegraphics[width=0.5\textwidth]{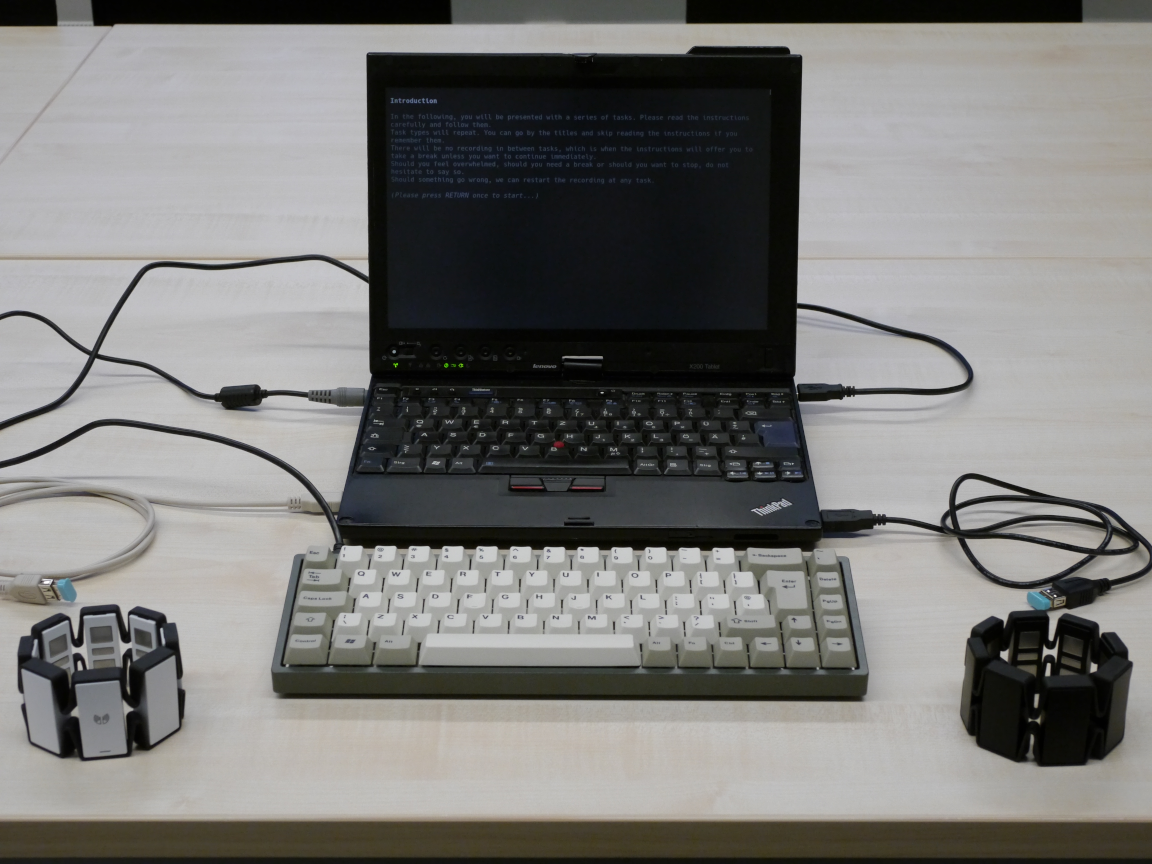}
\Description{A photograph of a laptop sitting on a table, in front of which the keyboard and the two Myo armbands are positioned, ready for a data recording. The laptop shows a black screen with the introductions for the participants displayed in white text.}
\caption{The setup for the data collection showing the two Myo armbands in the front.}
\label{fig:fig_setup}
\end{figure}

As part of our attempt to create realistic and diverse recording settings, the data study is carried out in multiple locations (including different offices, personal working spaces and some participant's living rooms), but the setup as shown in Fig.~\ref{fig:fig_setup} is the same in all cases.

The white Myo armband is worn on the left and the black one on the right arm, adjusted according to the official guidelines\footnote{Instructions by the manufacturer state that the Myo armband is supposed to be worn on the thickest part of the lower arm (below the elbow), with the USB port facing the wrist and the LED positioned on top of the arm (when stretched out).}.
The participants are encouraged to sit (or stand) and move as they usually do while typing.
The keyboard used for the data collection is a TADA68 with mechanical switches.
We support both the German and US international keyboard layout, in both cases as an ISO variant, asking the participants prior to the recording which layout they prefer.

\subsection{Study Design}

We design two distinctive data recording schemes, the first of which provides the \emph{training data} for the classifiers and the second of which provides the \emph{test data}.
The training and test data are recorded on completely separate occasions to account for biases associated with position changes of the armband.
In addition to that, both datasets contain data from participants who have only taken part in the training or test data recording respectively, in order to test the applicability of our approach on previously unknown persons.
Each recording scheme consists of multiple different tasks of different types, all but one of which prompt the user with groups of characters or words to copy.
During all recording sessions, we encourage the participants on multiple occasions to move or take a break and continue typing in a different position for diversification.

The training data scheme includes the following task types:
\begin{enumerate}
	\item \emph{Text}: This contains extracts copied from Wikipedia articles, selected such that they contain each key in consideration, as shown in Fig.~\ref{fig:keyboard_keys}, at least 10 times in total.
	\item \emph{Pangram}: Also copied from Wikipedia, each of these pangrams contain each letter of the alphabet (i.e.\ a subset of the keys in consideration) at least once.
	\item \emph{Random}: This contains pseudo-random groups of two, three or five different characters, generated by shuffling the set of characters corresponding to the keys in consideration such that each key appears exactly twice throughout this task. The rationale behind this task is to ensure that we get enough samples for each possible key, uniformly distributed within this task.
	\item \emph{Random (memorized)}: The same as the above with groups of two, three or four characters. However, this time the characters disappear once the participant starts typing, requiring them to remember the characters and preventing them from copying the characters one after the other. This design is inspired by the observation that users tend to type passwords in chunks of three to four quickly typed characters with a small pause in between \cite{song2001timing}.
	\item \emph{Game}: A console-based game which involves a small subset of the characters in consideration, included for diversification.
	This is the only task which does not require copying text.
\end{enumerate}

These tasks are meant to cover most of the types of input made on a physical keyboard in order to be able to infer text as well as passwords, the latter of which can either be closer to a random combination of characters or to text.
For all tasks except the game the participants see what they have typed and are able to correct mistakes using the \texttt{backspace} key.

For the test data scheme, we develop four different task types:
\begin{enumerate}
	\item \emph{Insecure}: Passwords which are randomly selected from 250 of the top 1000 passwords used \cite{website:most-common-pws}, e.g.\ \textit{iloveyou}.
	\item \emph{xkcd}: Passphrases inspired by xkcd \cite{website:xkcd-entropy} and consisting of six randomly chosen words from the Electronic Frontier Foundation list \cite{website:eff-long}, e.g.\ \textit{hatchery gratified drinking tiling precinct anywhere}.
	\item \emph{pwgen}: Passwords with a length of eight characters generated by pwgen, a password generator aimed at creating easy to type and easy to remember semi-random passwords, e.g.\ \textit{ei9Aemac}.
	\item \emph{Random}: Randomly generated passwords with a length of eight characters, e.g.\ \textit{p5nkmq'y}.
\end{enumerate}

These tasks are meant to cover both frequently used and insecure passwords, as well as passwords generated according to different guidelines such as securing a password by length while making it memorable (xkcd), using a password which is pseudo-random but easier to type and memorable (pwgen) and using a random password. Thus, we can evaluate and compare the performance of our classifiers on common but inherently different types of passwords of different strengths.
To better mimic normal password input behavior, the participants receive no feedback of what they type during entry (feedback is only given after hitting \texttt{return}) but they are allowed to correct mistakes using the \texttt{backspace} key.

We require the participants to type each password correctly four times (if they do make a mistake the entry does not count towards this) to familiarize them with the password and to simulate the entry of a known password to a certain extent. Note that correcting mistakes while typing is possible without triggering an additional repetition, as the check for correctness is only made after the password entry is confirmed by pressing \texttt{return}, as it would be the case for a real password entry.
We set this limit of four repetitions because early tests have shown that requiring a higher number of correct entries can be upsetting for the participants. For future studies, we suggest including repetitions of the same password spread across the whole recording to circumvent this problem.

The data study was approved by the ethics commission of TU Darmstadt (EK 24/2018). Vouchers were raffled off among the participants taking part in the longer recording of the training data.

\subsection{Evaluation}\label{sec:dataset}

A total of 37 volunteers took part in our data study (about one third female, two third male). Most were students or employees of the university, ranging in age between 18 and 74, with two thirds of the participants between the age of 25 to 34.

We distinguish between two typing styles: Touch typing and hybrid typing. The latter includes hunt-and-peck typists as well as typists with an incomplete touch typing training (e.g.\ who use less than ten fingers). The categorization is based on self-reporting in combination with our observations during the recordings. We discerned that even trained touch typists often diverge from the guidelines for the finger-to-key mapping, an observation already recorded in the literature \cite{feit2016we}.

Typing speeds reported by the participants range from about 86 to 500 keys per minute. Most participants use a keyboard on a daily basis.
The mean typing speed per person for the training data is 65 to 231 keys per minute. For the test data, it is 125 to 235 keys per minute. This is significantly lower than the reported typing speed, which is mostly due to the random character tasks for which the average typing speed is 71 keys per minute across all participants (compared to 241 keys per minute for the text-based tasks). Other influences reducing the typing speed may be the unfamiliar keyboard or the reduced look-ahead due to the limited number of characters presented to the participant at a time (one sentence or pangram for the text-based tasks, one chunk of characters for the random tasks and one password for each task type in the test data scheme). According to prior research, look-ahead and motor preparation generally have a significant effect on typing speed \cite{salthouse1984effects, feit2016we}. We also observe a difference in the average typing speed for touch typists with 161 keys per minute and hybrid typists with 134 keys per minute for the training data with an even larger difference for the test data of 184 keys per minute for touch typists and 164 keys per minute for hybrid typists.

\begin{figure}
\centering
\subfigure{\includegraphics{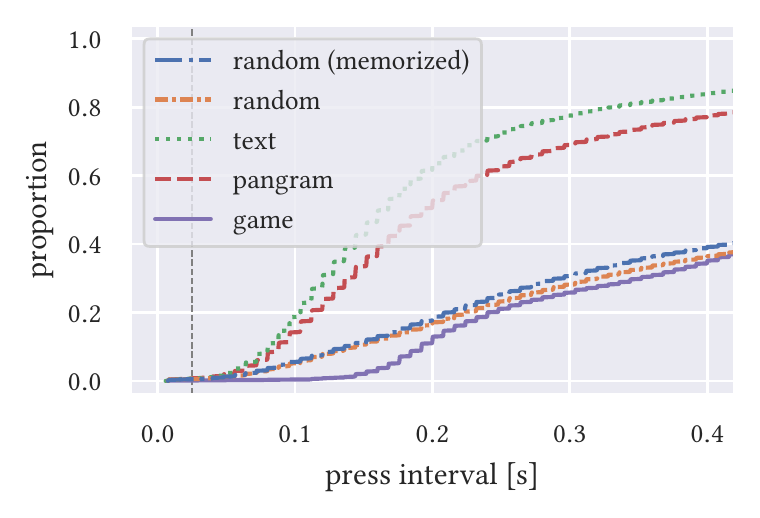}}
\Description{A cumulative distribution of the time intervals in seconds between one keypress event and the next ranging from 0 to 0.4 seconds. The curves for the text and pangram tasks run close to each other, the pangram depicting a slightly larger proportion of keys with larger intervals. Both show a quickly rising curve after about \unit{0.05}{\second}, passing \unit{60}{\%} of the intervals for text and \unit{50}{\%} for pangram after \unit{0.2}{\second}. After that the curves rise slower, including around \unit{80}{\%} of the intervals at \unit{0.4}{\second}. The curves for the two random tasks are even closer to each other. The curves do not depict a steep rise in the beginning such as the text-based tasks but rise slowly starting after about \unit{0.05}{\second} and including about \unit{40}{\%} of the intervals at \unit{0.4}{\second}. Finally, the game task shows a similar curve shape to the text-based tasks but start only after about \unit{0.15}{\second} and stay close to the random tasks after a small steep rise which is followed by a slower rising curve, also including about \unit{40}{\%} of the intervals at \unit{0.4}{\second}.}
\subfigure{\includegraphics{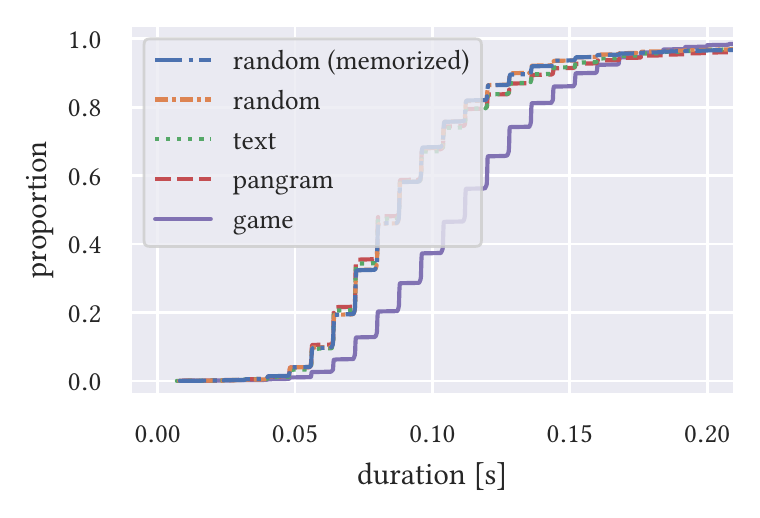}}
\Description{A cumulative distribution of the duration of one keystroke in seconds ranging from 0 to 0.2 seconds. The curves for the all tasks run close to each other except for the game task which shows a longer duration for keystrokes. All depict a steep curve including almost all keystroke duration between \unit{0.05}{\second} and \unit{0.15}{\second} with the game curve being shifted slightly to the right.}
\caption{Cumulative distribution of press-press intervals for different task types (left) and keystroke duration, i.e.\ times between press and release of a key (right), in the training data. The vertical line on the left marks the minimum peak distance for the peak detection.}
\label{fig:pp_intervals}
\end{figure}

Fig.~\ref{fig:pp_intervals} shows the difference in typing patterns between the different task types in the training data regarding the press-press intervals between the press events of successive keystrokes (left) and the duration of the keystrokes from press to release (right). In particular, it shows the diversification gained by the game task. We added the random (memorized) task hoping to get a more fluent typing pattern if the participants had to remember the characters, but the results show no significant difference in typing speed between the two random task types. This is most likely due to some participants having difficulty remembering the characters while most participants did not copy them character by character even if they remained visible. In fact, disabling the feedback of what the participants type seems to have a larger effect in this case, as many participants tend to check if they copied the characters correctly, which disrupts the typing flow.

The \emph{class skew} refers to the number of samples of a given class as compared to the number of samples of the other classes representing the imbalance inherent to the dataset. For a keystroke detection on keypress events, the positive class refers to the single point in time of an actual keypress and the negative class covers every step in time that does not contain a keypress. In our training data, the mean class skew is \unit{98.7}{\%} for the negative class compared to the positive class. For the test data, the mean class skew is \unit{98.3}{\%}. Thus, the dataset is highly imbalanced towards the negative class.
Typing speed and class skew negatively correlate with each other, but the changes are insignificant given the overall bias.
The mean class skew for the fastest typist is \unit{99.5}{\%} on the training data and \unit{98.8}{\%} on the test data.
For the slowest typist it is \unit{98.1}{\%} on the training data and \unit{97.7}{\%} on the test data.
The class skew for a key identification depends on the number of samples given for each key. In our dataset, this is uniform for the random tasks but mostly underlies language characteristics for the text-based tasks. \texttt{Space} is the most frequent key.

Each participant of the data study took part in either the test or the training data recording scheme or in both of them, resulting in a total of 29 training and 17 test data recording sessions, six of which were recorded with the same participants for training as well as test data. Furthermore, the training data contains three additional recordings by two participants. Thus, we have a total of 37 participants for a total of 46 recording sessions.

For the training data scheme, we include a total of 24 recordings of separate tasks (one game, five pangram and six of every other task type) with more than 8000 keys recorded from each participant on average. The training data includes 261\,962 keystrokes in total.
For the test data scheme, we have 12 tasks (three for each type, each including two passwords repeated at least four times). Two thirds of the tasks include the same passwords for multiple participants, one third includes a random selection of passwords different for each participant. The dataset includes 48\,604 keystrokes in total.
On average, a single data recording took one hour for the training data and 30 minutes for the test data scheme. This includes small breaks and movements made by the participant during a task recording. In total, this amounts to 310\,566 keystrokes and about \unit{34}{\hour} of typing.

\begin{figure}
\centering
\includegraphics{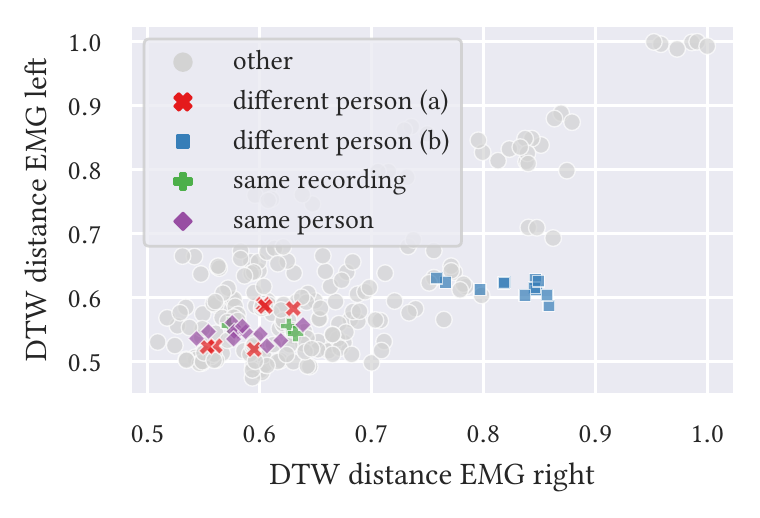}
\Description{A scatter plot showing the \ac{DTW} distances of multiple recordings of the same word for the left and right arm as points between 0.5 and 1. The recordings of the same participant partially overlap with the recordings of that participant taken in another recording session, as well as with those of a different participant, showing similar distances across recordings but also across participants. However, not all participants display close or overlapping recordings, as is shown for one participant marked in blue, whose recordings show a significant difference in the distances of the right hand to the aforementioned participants. The data of all other participants is shown grey in the background, most of these dots are close to each other with a couple of outliers.}
\caption{The \ac{DTW} distance between different recordings in the training data of the word \emph{Quetzalcoatl} of the same participant recorded in the same (green) and a different recording session (violet) and of different participants (red and blue). The recordings made during the same session are closest to each other but the recordings made in a different session partly overlap those by the participant marked in red. On the other hand, these recordings do not overlap at all with those of the participant marked in blue. These participant's data is highlighted as an example, the other recordings are shown in grey.}
\label{fig:within_between_user_quetzalcoatl}
\end{figure}

In order to compare the feasibility of a within-subject and a between-subject approach, we apply \ac{DTW} to repeating sequences representing the same, correctly typed words within the training data recordings and compare the resulting distances across different recordings. Fig.~\ref{fig:within_between_user_quetzalcoatl} shows the results for one word which appears up to five times in the tasks generated for the German keyboard layout during one recording session. It highlights that the distances between two recordings of the same session are more similar than those between different recording sessions, which show a similar distance to the recordings of different participants. This implies that a within-subject approach may only work well if the model is retrained whenever the Myo armband is taken off and put on again, which is impractical for an unsupervised approach and infeasible for a supervised approach. Thus, a between-subject approach seems best even when focusing on single users.

\section{Classification}\label{sec:implementation}
% goal (between-subject, supervised learning)
We formulate the problem of inferring keystrokes on sensor data (recorded within a typing scenario) as a two-step \emph{supervised classification} in a between-subject scenario, enabling us to infer keystrokes belonging to previously unknown persons.
% keystroke detection (context: samples in which typing occurs -- being in the context of typing)
Our first step is the keystroke detection, a binary classifier predicting whether a sample belongs to a keystroke or not by encoding the keypress event.
% press vs release encoding, vs state encoding
Due to the high class skew inherent to this encoding, as shown in Section~\ref{sec:dataset}, we considered predicting the key state instead, i.e.\ each moment in time in which a key is pressed.
However, this approach has its own disadvantages, most notably that it would require a multi-label approach to predict overlapping keystrokes, e.g.\ as required for modifier keys like the shift key.

% key identification (context: having a sample of keystroke)
The second step, the key identification, is a multi-class classifier which predicts the actual physical key on the samples predicted to belong to a keypress.
We have 52 classes in total representing the keys on the physical keyboard as depicted in Fig.~\ref{fig:keyboard_keys}.

% explain/introduce segmentation
The input to both classifiers is a segment of sensor data with a size of \unit{150}{\milli\second} before and \unit{100}{\milli\second} after the sample, which coincides with the actual keypress.
The overall segment size of \unit{250}{\milli\second} is similar to the one used in \cite{Zhang2017} and the size and position of the keypress within the segment are supported by the findings of \cite{Dennerlein1998}.
They report that bursts of \ac{EMG} activity in the forearm during a keystroke occur about \unit{150}{\milli\second} before and about \unit{50}{\milli\second} after bottoming out the key switch, which is very close to the time point of the key switch actuation point.
Similarly, they report that the downward movement of the raised finger starts about \unit{83}{\milli\second} before fully depressing the key switch.

% explain that no model has some sort of memory or whatsoever
As a consequence of our approach, both classifiers are state-less and therefore classify segments without considering past or future predictions.
This is necessary for the key identification as we are aiming at classifying unstructured text like passwords and therefore must not consider the likelihood of certain character sequences.
However, for the output of our keystroke detection, we apply a post-processing step, which takes the timing of keystrokes into account to eliminate false positives by enforcing a minimum time distance between consecutive keystrokes.

% overview of our classification pipeline
In summary, our proposed system consists of a simple pre-processing of the sensor data, which is described in the following, followed by the binary classification for keystroke detection, the post-processing of the binary results, and finally the multi-class classification for key identification.
Every design choice, including the hyperparameter optimization of the classifiers, as well as the choice of the parameters for the pre- and post-processing steps are based on the training data.
The source code for training the proposed models and reproducing our final results is available as free software at \url{https://github.com/seemoo-lab/myo-keylogging}.

\subsection{Pre-processing}
Each armband contains eight \ac{EMG} sensors sampled at \unit{200}{\hertz} and an \ac{IMU} containing an accelerometer, gyroscope and a magnetometer sampled at \unit{50}{\hertz}.
For this work, we do not consider the magnetometer as the provided rotation quaternion is deemed unsuitable for our task.
Since the armband does not provide timestamps for sensor values, we determine those by estimating the sampling frequency for each modality separately.

We tested filtering the \ac{EMG} sensor data by applying a second-order Butterworth high-pass filter with a cutoff frequency of \unit{20}{\hertz} following best practices \cite{DeLuca2010}.
However, we decided to exclude this filter as we observed a performance degradation.
Similarly, the performance degraded slightly when changing the cutoff frequency to \unit{10}{\hertz} as in similar work with the Myo armband \cite{Fu2020}, as well as when testing a second-order \ac{IIR} notch filter at \unit{50}{\hertz} ($Q=25$) to remove the \ac{PLI}.
Thus, we do not apply any filtering on the data.
Instead, we rely on the neural networks to discern relevant parts of the raw data, enabling an end-to-end solution.

\begin{figure}
  \centering
  \subfigure{\includegraphics{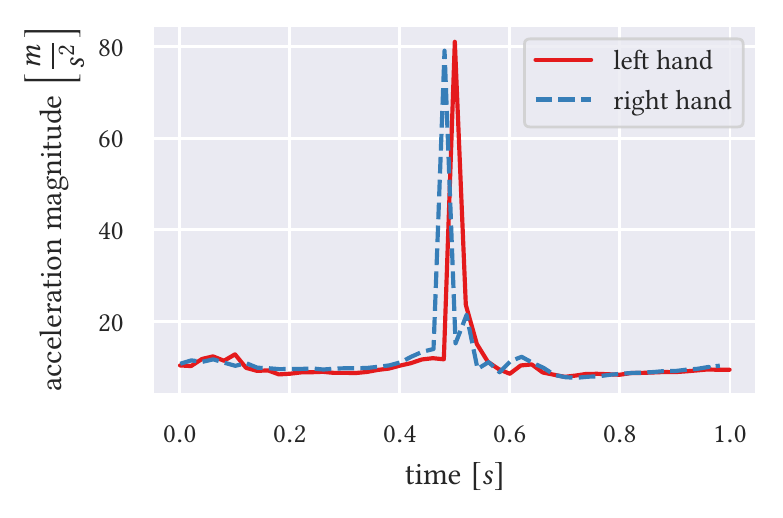}}
  \Description{A line plot showing a sample of a clap represented in the accelerometer data with a magnitude of about \unit{80}{\metre\per\second\squared} for both the left and the right hand, both peaks of which align at around \unit{0.5}{\second}. The peaks are clearly distinguished from the data around them which is stable at around \unit{10}{\metre\per\second\squared}.}
  \subfigure{\includegraphics{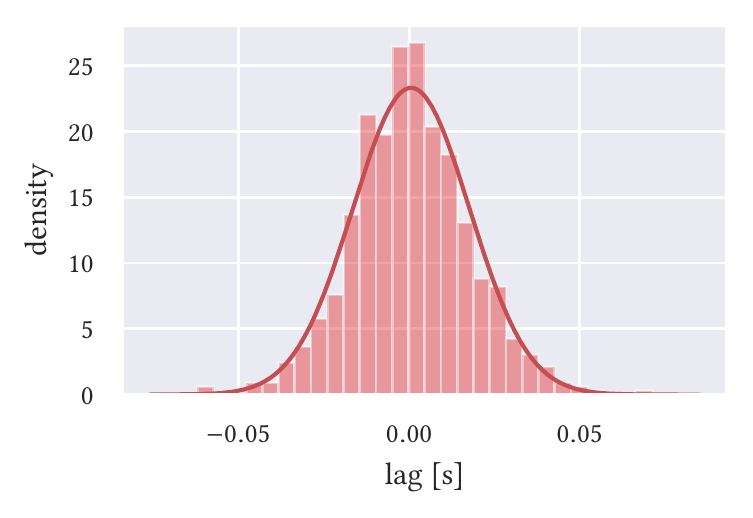}}
  \Description{A bell-like histogram centered at 0 and fit with a Gaussian curve with most values between \unit{-0.05}{second} and \unit{0.05}{second}.}
  \caption{Left shows an example of the accelerometer magnitude values of both Myo armbands during the joint high acceleration event (clapping the hands). Right shows the distribution of the peak-to-peak time differences between those values for all training samples. The curve in the right picture is a normal distribution fitted to the data points.}
  \label{fig:time_lag_dist}
\end{figure}

We fuse the sensor data of a single armband by resampling the \ac{IMU} data to match the higher \ac{EMG} sample rate.
An offset between the armbands' sensor values is calculated by recording a joint high acceleration event, namely the clapping of both hands as exemplarily shown on the left in Fig.~\ref{fig:time_lag_dist}.
This latency value is then used to align the data of both armbands, followed by a resampling of both data streams to a common sampling rate.
Note that this alignment cannot be applied when considering a realistic attack scenario.
Though the absolute lag between both devices is low, being considerably less than \unit{50}{\milli\second} for almost every sample, as can be seen in the histogram on the right in Fig.~\ref{fig:time_lag_dist}.
We assume that in practice, this makes no difference, as we did not observe a performance degradation when disabling this sensor alignment.

\subsection{Neural Networks}

% binary vs multiclass model configuration
We train four different neural networks, each of which has shown promising results when applied to problems that are similar to ours in one way or another while being based on a different approach than the other models. All models are trained once for binary and once for multi-class classification using the training data in our dataset.
All classifiers are designed as end-to-end neural networks:
We skip feature engineering, training the classifiers on raw data and thus relying on the neural networks to find good feature representations relevant for the given task.
However, in order to connect the binary with the multi-class classifiers, we require an intermediate step of post-processing the binary output with a peak detector in order to remove candidates of predictions which likely belong to the same keystroke.

Each model contains a final \emph{output layer}, linearly mapping the output of the previous stages onto either a single neuron (binary classification) or on a number of neurons matching the number of required classes (multi-class).
For binary classification, we apply the sigmoid function on the output layer, whereas for multi-class classification we use the softmax function.
Through hyperparameter optimization on both types of classification problems, we did not find notable differences justifying different sets of hyperparameters.
Note that simpler prototypes based on \ac{CNN} showed greater specialization towards a single type of classification when comparing the performance with a single hyperparameter set, though these attempts were consistently outperformed by the four models described below. More details for the hyperparameters of each model are given in Appendix~\ref{sec:appendixclassifiers}.

\subsubsection{TSC ResNet11}
% intro, references and motivation
This neural network is a close recreation of an 11-layer ResNet \cite{he2016deep} adaption proposed \cite{wang2017} and reviewed \cite{fawaz2019} as a strong off-the-shelf choice in univariate, as well as multivariate \ac{TSC} and is, therefore, a natural fit to be applied to our problem.
Our main change to the original architecture is the introduction of so-called grouped convolutions \cite{ioannou2017}, essentially providing independent paths through the network for each of the 28 sensor input channels.
In our experience, this single architectural change boosted the performance significantly.

\subsubsection{CWT ResNet18}
This neural network is a combination of a batch normalization layer, followed by a \ac{CWT} and an 18-layer ResNet network \cite{he2016deep} with full pre-activation residual units \cite{He2016} inspired by a similar approach for hand gesture recognition based on the \ac{EMG} sensor data from a Myo armband \cite{cote2019deep}.
The basic idea of this approach is to use the \ac{CWT} ($\delta_j = 0.125$, $\delta_t = 0.005$) with a Morlet wavelet ($\omega_0 = 6$ as suggested in \cite{torrence1998}) to transform the one-dimensional time series representation of each individual sensor channel into a two-dimensional representation, namely the wavelet power spectrum \cite{torrence1998}, and apply a popular image recognition model like ResNet to perform the actual classification.
Our main change to the basic ResNet architecture is the removal of the early max pooling layer, which could be explained by having a significantly smaller input size:
The wavelet power spectrum for each segment consists of 38x50 data points compared to the original input of a 224x224 pixel-sized image \cite{he2016deep}.

\subsubsection{CRNN}
% intro, references and motivation
The \ac{CRNN} is a neural network architecture consisting of one or more convolutional layers followed by one or more recurrent layers.
This particular network is inspired by works in the related field of \ac{HAR} \cite{burns2018, ordonez2016}, dealing with a related task of classifying multivariate sensor data.
The main difference compared to our task of inferring keystrokes is the relatively long time span of human activity classes compared to single keystrokes.
% architecture
Our final architecture consists of a batch normalization layer, a single convolution layer with grouped convolutions, followed by a \ac{ReLU} function and batch normalization, and a series of two \ac{LSTM} layers.

\subsubsection{TSC WaveNet}
% intro, references and motivation
This neural network is inspired by the WaveNet architecture \cite{Oord2016} and the Gated PixelCNN architecture using dilated convolutions \cite{Oord2016pixel}.
Although being designed for audio synthesis, the authors already suggested using the WaveNet architecture for classification tasks.
To cope with the multivariate nature of our sensor data, we introduce grouped convolutions like explained earlier for the \ac{CRNN} and \ac{TSC} ResNet11 architectures.
Aside from this change, our implementation closely follows the structure of the original work with two residual block layers.

\subsection{Training Process}

% 3-fold CV HPO
% grouped k-fold for between-subject (temporal split for within-subject)
% f1-score for ranking hyperparams
We perform a 3-fold cross-validated hyperparameter random search on the training data in order to assess the mean performance and stability of a hyperparameter set across all folds.
The folds are created such that the data from a single participant is only present in a single fold in order to find hyperparameters which generalize well between different persons.
The f1 score is used to rank the hyperparameter sets.

% Adam or RMSprop with cross-entropy loss
% train-valid split
We use Adam ($\beta_{1} = 0.9, \beta_{2}=0.999$) as the gradient descent optimizer for the TSC ResNet11, CWT+ResNet18 and TSC WaveNet models and RMSprop ($\alpha=0.99$) for the CRNN as indicated in Table~\ref{table:optimizer} together with the cross-entropy loss.
During training, the last \unit{20}{\%} of the training data are used as validation data to apply early stopping.

\begin{table}
  \caption{Optimizer configuration and batch size per neural network architecture.}
  \label{table:optimizer}
  \begin{tabular}{ccccc}
    \toprule
    Model & Optimizer & Learning Rate & L2 Weight Decay & Batch Size\\
    \midrule
    TSC ResNet11 & Adam    & 0.001  & 0       & 16  \\
    CWT+ResNet18 & Adam    & 0.0001 & 0.00001 & 512 \\
    CRNN         & RMSprop & 0.001  & 0.001   & 128 \\
    TSC WaveNet  & Adam    & 0.005  & 0       & 128 \\
    \bottomrule
  \end{tabular}
\end{table}

% data imbalance binary and multi-class
% binary: random undersampling per epoch
% multi-class: class weights
Our data exhibits two types of class skew, one relevant for the binary and one for the multi-class classification as described in Section~\ref{sec:dataset}.
When training a binary classifier, we apply random subsampling of the majority class (non-press events) after each training epoch to compensate for the comparatively small number of keypress events.
For multi-class classification, we counter the imbalance of typing different keys a different amount of time by using class weights which are inversely proportional to their occurrences in the training data.

\subsection{Post-processing}\label{sec:postprocessing}

% 1. peak detector --> reduce broad prediction to single peaks
% 2. min height --> Reuse the (arbitrary) default threshold of 0.5 to determine peaks
% 3. min distance of 5 samples (25 ms) --> exploit the temporal dependency of consecutive keystrokes to reduce the number of false positives (this serves as kind of sanity check)
% 4. min prominence of 0.05 --> smooth the prediction to reduce false positives due to wiggles in the models output

% 1. reduce to rather uncertain/broad output to single peaks
% 2. discern close peaks vs suppress wiggles (false positives)

Even though our classifiers are trained on keypress events, their prediction of a keypress usually spans multiple timestamps.
This is because each prediction made by our classifiers is independent of the prediction on neighboring samples. However, two keypresses rarely appear in two consecutive samples.
We apply a simple peak detector on the output of the binary classifiers to refine the output by taking this temporal dependency between consecutive keystrokes into account.

We manually define a minimum peak distance of \unit{25}{\milli\second}, which is smaller than \unit{99}{\%} of all press-press intervals between keystrokes within the training data as shown on the left in Fig.~\ref{fig:pp_intervals}.
To further reduce the number of false positives, we require a minimum peak height of 0.5 with a minimum prominence of 0.05.

However, the prediction values resulting from this peak detection do not necessarily match the ground truth exactly but may be shifted by a couple of samples.
We argue that a shift by a few samples should not have an effect on the overall results or the multi-class prediction, similar to the temporal tolerance applied in time series segmentation \cite{gensler2014} and anomaly detection \cite{lavin2015, tatbul2018} in which the presence of a prediction is more important than an exact match.
Thus, we apply a custom metric that is tolerant given a \emph{temporal tolerance}, i.e.\ a certain maximum shift in time, between the prediction of the classifier and the encoded truth.
We formulate the problem of matching the ground truth keystrokes to the ones in the prediction as a linear assignment problem with the cost linearly increasing with the lag between the predicted and actual keystroke.
Additionally, we assign an infinitely high cost to pairs with a distance larger than a given temporal tolerance, in order to avoid matching them.
This allows us to create a temporally tolerant confusion matrix as described in Equations~\ref{eq:one}-\ref{eq:four}, where the \texttt{positives} are samples of keypress events and the \texttt{negatives} are samples without keypress events.
\begin{align}
\label{eq:one}
\texttt{true\_positives} &= \texttt{matched\_positives}\\
\label{eq:two}
\texttt{false\_positives} &= \texttt{number\_of\_predicted\_positives} - \texttt{matched\_positives}\\
\label{eq:three}
\texttt{false\_negatives} &= \texttt{number\_of\_actual\_positives} - \texttt{matched\_positives}\\
\label{eq:four}
\texttt{true\_negatives} &= \texttt{number\_of\_predicted\_negatives} - \texttt{false\_negatives}
\end{align}

Given the values of this confusion matrix, we evaluate the results on standard metrics to gain a performance estimate of the peak detection.

%%%%%%%%%%%%%%%%%%%%%%%%%%%%%%%%%%%%%%%%%% Results %%%%%%%%%%%%%%%%%%%%%%%%%%%%%%%%%%%%%%%%%%%%%%%%%

\subsection{Keystroke Detection Results}

In the following, we present the performance results of evaluating our trained classifiers on the test data of our dataset described in Section~\ref{sec:dataset}.
In this section, we evaluate the keystroke detection, the key identification results are given in Section~\ref{sec:results_id}.

Table~\ref{tab:results_detection} shows the average prediction performance of each binary classifier on the entire test data, i.e.\ evaluated on all participants and all password types, without applying peak detection. The performance of the classifiers is similar with a mean balanced accuracy of about \unit{74-76}{\%} except for the TSC WaveNet which performs worse than the others with a mean balanced accuracy of only \unit{70.7}{\%}. CRNN and TSC ResNet11 perform best, the former showing better results on the mean balanced accuracy (\unit{76.1}{\%}) and the latter showing better results on the mean f1 score (\unit{10.5}{\%}).

\begin{figure}
\centering
\includegraphics{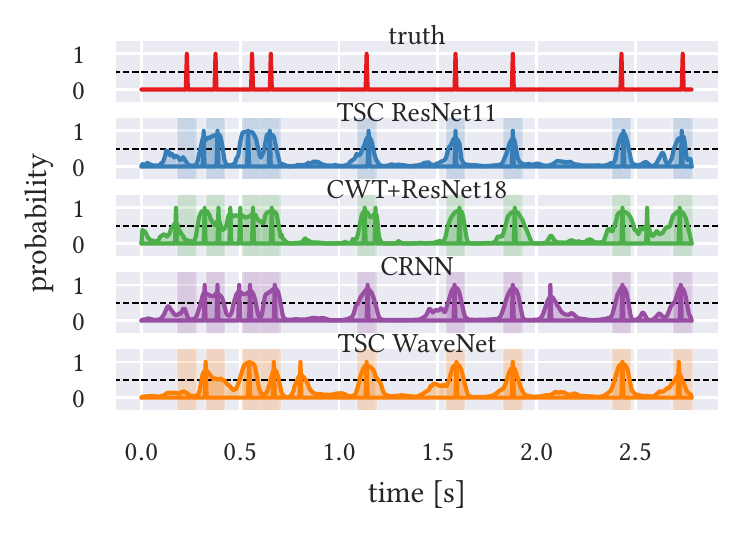}
\Description{A sample plot showing subplots of the truth and the predictions of all classifiers on one password as probabilities between 0 and 1 with a threshold of 0.5 (probabilities above this threshold are regarded as keystrokes). The truth is depicted as single peaks of probability 1 in the topmost subplot, a temporal tolerance of \unit{0.05}{\second} around these peaks are included as semi-transparent boxes in all prediction subplots for comparison with the predictions made by the estimator. The prediction of each classifier includes both the probability, rising and falling around predicted keystrokes, and the peak predictions shown as single peaks of a probability of 1 like the truth. In this particular sample the TSC ResNet11 comes closest to the truth regarding the peaks, followed by the TSC WaveNet which has the correct number of peaks but some are shifted farther than the temporal tolerance of \unit{0.05}{\second}. The CRNN shows two false positives and the CWT+ResNet18 sometimes predicts multiple peaks for one true keystroke resulting in multiple false positives. None of the classifiers has predicted less keypresses than are present in the truth. CWT+ResNet18 also shows the broadest predictions. All models show low probabilities for most of the wider gaps between keystrokes.}
\caption{Sample truth and prediction values for the password \textit{wah/Ri2t} typed on a German keyboard layout with the \texttt{shift} key covering two consecutive keys (\textit{/R}). Both the prediction probabilities and the peak predictions of each classifier are given. The semi-transparent background marks a temporal tolerance of \unit{50}{\milli\second} before and after the truth.}
\label{fig:binary_prediction_example}
\end{figure}

The comparatively weak f1 score of about \unit{9-10}{\%} and precision of about \unit{5-6}{\%} can be explained by the high class skew and the observation that none of the trained classifiers produces a point-wise prediction to match the given truth. Instead, they show a wider peak of increasing and decreasing probabilities as shown in the example result in Fig.~\ref{fig:binary_prediction_example}. This example of a keystroke detection on a pwgen password also allows some observations on the behavior of the classifiers. For example, a small distance between two keypresses often results in two merging peaks while keystrokes further apart from each other show a pronounced peak for the majority of the predictions. Peaks corresponding to false positives are often smaller than those corresponding to true positives. Similarly, false negatives are often still visible as a small peak.

\begin{table}
\caption{Mean and Standard Deviation of Keystroke Detection Results}
\label{tab:results_detection}
\begin{tabular}{ccccc}
\toprule
\textbf{Classifier Type} & \textbf{Balanced Acc. (\%)} & \textbf{F1 Score (\%)} & \textbf{Precision (\%)} & \textbf{Recall (\%)}\\
\midrule
TSC ResNet11	& 74.6 (9.1) &  10.5 (2.3) &  5.7 (1.3)  &  76.3 (18.1) \\
CWT+ResNet18	& 74.0 (9.8) &  9.1 (2.4)  &  4.9 (1.6)  &  81.1 (20.2) \\
CRNN			& 76.1 (9.6) &  10.1 (2.6) &  5.4 (1.4)  &  82.1 (18.9) \\
TSC WaveNet		& 70.7 (9.5) &  8.8 (2.8)  &  4.8 (1.6)  &  74.5 (23.2) \\
\bottomrule
\end{tabular}
\end{table}

\begin{figure}
\centering
\subfigure{\includegraphics{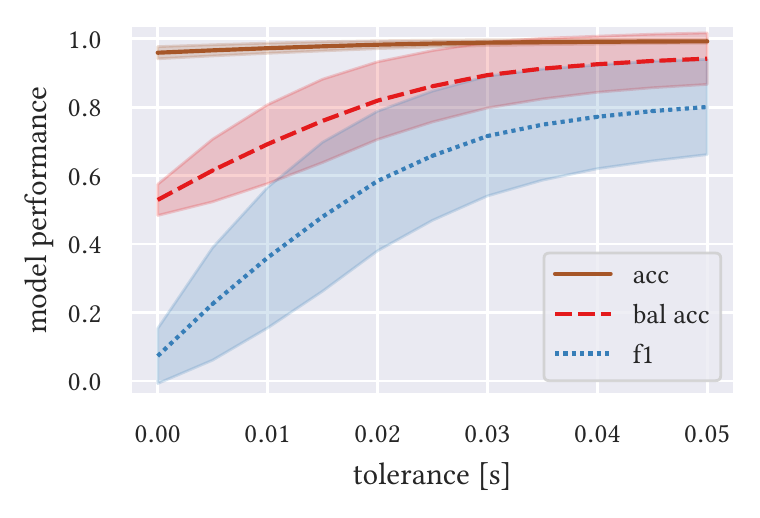}}
\Description{A line plot showing the results of the mean accuracy, the mean balanced accuracy and the mean f1 score for different temporal tolerances around a peak detection rising in a slowly flattening curve from zero tolerance and a result equal to guessing for each metric to a tolerance of 0.05 seconds and a mean accuracy of about \unit{99}{\%}, a mean balanced accuracy of about \unit{94}{\%} and a mean f1 score of about \unit{80}{\%}. The mean accuracy hardly changes, as the dataset is highly imbalanced, resulting in a mean accuracy of \unit{96}{\%} for a temporal tolerance of zero. The standard deviation ranges from \unit{2}{\%} to \unit{1}{\%} for the accuracy, from \unit{5}{\%} to \unit{8}{\%} for the balanced accuracy and from \unit{8}{\%} to \unit{14}{\%} for the f1 score.}
\subfigure{\includegraphics{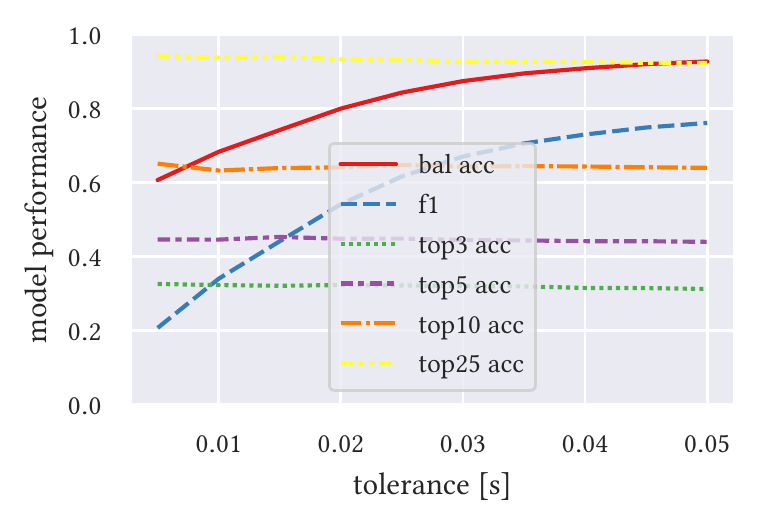}}
\Description{A line plot showing the binary results of the mean balanced accuracy and the mean f1 score for different temporal tolerances around a peak detection rising in a slowly flattening curve from a tolerance of 0.005 and a result of a mean balanced accuracy of about \unit{61}{\%} and a mean f1 score of about \unit{23}{\%} for each metric to a tolerance of 0.05 seconds and a mean balanced accuracy of about \unit{94}{\%} and a mean f1 score of about \unit{80}{\%}. Along with this, the multi-class results are depicted with a mean top-3 key accuracy of about \unit{32}{\%}, a mean top-5 key accuracy of about \unit{44}{\%}, a mean top-10 key accuracy of about \unit{64}{\%} and a mean top-25 key accuracy of about \unit{92}{\%}. These accuracies remain stable over time.}
\caption{The CRNN (left) and CWT+ResNet18 result (right) of the mean temporary f1 score and mean temporary balanced accuracy on the peak predictions of each classifier given different temporal tolerances along with the mean accuracy (left) and along with the mean top-n key accuracies (right). The standard deviation is omitted for better visibility on the right. The x axis of both plots depicts the temporal tolerance, i.e.\ the accepted distances around each keypress encoded in the truth within which a prediction is considered as a correct prediction. A distance of zero equals an evaluation of the peak detection based on the standard metrics without temporal tolerance.}
\label{fig:binary_temp_predictions}
\end{figure}

Applying a peak detection to receive the point-wise predictions shown in Fig.~\ref{fig:binary_prediction_example} poses the problem that most of these predictions do not match the exact timestamps of the keypress event encoded in the truth as discussed in Section~\ref{sec:postprocessing}. This can be seen for a distance of zero on the left in Fig.~\ref{fig:binary_temp_predictions} which equals the results of a classifier that almost always predicts the majority class.
Thus, to give an estimate of the performance of the binary classifiers with a peak detection, we compare the results on different temporal tolerances as described in Section~\ref{sec:postprocessing}. The left part of Fig.~\ref{fig:binary_temp_predictions} shows the results of this evaluation with both the performance and variance increasing for a temporal tolerance of zero (i.e.\ the prediction must match the truth exactly) to \unit{50}{\milli\second} the latter of which is depicted as a semi-transparent window in Fig.~\ref{fig:binary_prediction_example}. For a window of \unit{50}{\milli\second}, the CRNN achieves a mean balanced accuracy of \unit{94.2}{\%} and a mean f1 score of \unit{80.0}{\%}. We observe that the lags between the true and the predicted keypress events are normally distributed. For a distance of \unit{50}{\milli\second} to the encoded true keypress the mean lag is \unit{-4.1}{\milli\second} for the CRNN and \unit{-2.6}{\milli\second} for the TSC ResNet11 with a standard deviation of \unit{21.55}{\milli\second} and \unit{20.65}{\milli\second} respectively. Thus, most keypresses are predicted slightly before they occur according to the ground truth.

\begin{table}
\caption{Mean and Standard Deviation of Keystroke Detection Results for Touch and Hybrid Typists (CRNN and TSC ResNet11)}
\label{tab:results_typist_type}
\begin{center}
\begin{tabular}{ccccc}
\toprule
\textbf{Typist} & \multicolumn{2}{c}{\textbf{Balanced Accuracy (\%)}} & \multicolumn{2}{c}{\textbf{F1 Score (\%)}}\\
\textbf{} & \textbf{\textit{TSC ResNet11}} & \textbf{\textit{CRNN}} & \textbf{\textit{TSC ResNet11}} & \textbf{\textit{CRNN}} \\
\midrule
touch	& 71.3 (8.6) &  73.1 (9.6) &  10.3 (2.7) &  10.3 (2.8) \\
hybrid	& 77.5 (8.6) &  78.7 (8.8) &  10.6 (2.0) &  9.9 (2.4)  \\
\bottomrule
\end{tabular}
\end{center}
\end{table}

\subsubsection{Typing Style}
On average, the results of hybrid typists compared to touch typists are better regarding the balanced accuracy but worse or almost equal regarding the f1 score as shown in Table~\ref{tab:results_typist_type}. This is likely due to the typing speed difference between the two groups, given that in our dataset touch typists type faster on average than hybrid typists, as described in Section~\ref{sec:dataset}, and the typing speed correlates negatively with the performance of the binary models, i.e.\ the mean balanced accuracy drops for higher typing speeds, as shown on the left in Fig.~\ref{fig:performance_speed_comparison}. This could be explained by a lower number of positive predictions (both correct and false) due to an increased chance of the classifier to completely miss a keystroke, which is more likely to happen for faster typists. Also, multiple keys in rapid succession leave less samples between keys for false positives. On the other hand, predictions of multiple quickly typed keystrokes will likely merge, such as the first four keys in Fig.~\ref{fig:binary_prediction_example}.

\subsubsection{Password Types}
\begin{figure}
\centering
\subfigure{\includegraphics{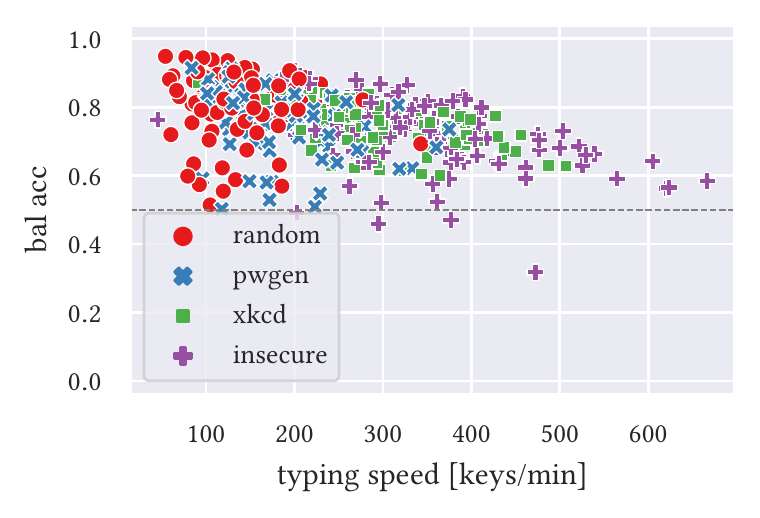}}
\Description{A scatter plot showing the distribution of the mean balanced accuracy of the CRNN model over the typing speed of the participants for each individual password recording. The colors indicate the type of password. The distribution of the dots shows a negative correlation, ranging from about \unit{95}{\%} for slower typing with about 50 keys per minute to around \unit{60}{\%} for faster typing of about 500-600 keys per minute. The random password samples are clustered at the far left of the speed range, the pwgen password following also to the left, both ranging to a maximum of 400 keys per minute for a few samples. The xkcd and insecure password samples mostly start at about 200 keys per minute and range to the far right of the speed range with the insecure samples showing the most variation in performance. A horizontal line is drawn to show the probability for guessing with most dots above this line.}
\subfigure{\includegraphics{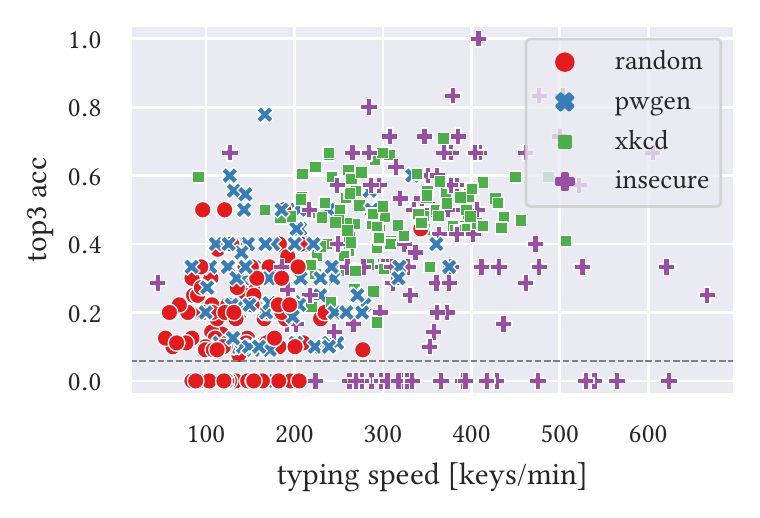}}
\Description{A scatter plot showing the distribution of the mean top-3 key accuracy of the \ac{CWT}+ResNet18 model over the typing speed of the participants for each individual password recording. The colors indicate the type of password. The distribution of the dots shows a cloud with mean top-3 key accuracies of \unit{10-60}{\%} with some samples at \unit{0}{\%}. The random password samples are clustered at the far left of the speed range, the pwgen password following also to the left, both ranging to a maximum of 400 keys per minute for a few samples. The xkcd and insecure password samples mostly start at about 200 keys per minute and range to the far right of the speed range with the insecure samples showing the most variation in performance. Insecure passwords also show the most samples at \unit{0}{\%} mean top-3 key accuracy with pwgen and xkcd showing none and xkcd passwords showing very consistent results mostly between \unit{50-60}{\%}. A horizontal line is drawn to show the probability for guessing with most dots above this line.}
\caption{The mean balanced accuracy for the keystroke detection evaluated on the CRNN model (left) and mean top-3 key accuracy for the key identification evaluated on the \ac{CWT}+ResNet18 model (right) for each password recording compared to the current typing speed of the respective participant. The horizontal line marks the probability of guessing.}
\label{fig:performance_speed_comparison}
\end{figure}

A similar trend for performance applies for the different password types shown in Table~\ref{tab:results_password_type} and on the left in Fig.~\ref{fig:performance_speed_comparison}, as (pseudo)random passwords are typed in a slower speed than those resembling text as mentioned in Section~\ref{sec:dataset}. The slight decrease in the mean balanced accuracy for pwgen passwords shows that these are slightly easier to type than random passwords but harder than passwords resembling text. Interestingly, the insecure passwords show the widest performance spread, likely because most of them are short and rapidly typed.

\begin{table}
\caption{Mean and Standard Deviation of Keystroke Detection Results per Password Type (CRNN and TSC ResNet11)}
\label{tab:results_password_type}
\begin{tabular}{ccccc}
\toprule
\textbf{Password} & \multicolumn{2}{c}{\textbf{Balanced Accuracy (\%)}} & \multicolumn{2}{c}{\textbf{F1 Score (\%)}}\\
\textbf{Type} & \textbf{\textit{TSC ResNet11}} & \textbf{\textit{CRNN}} & \textbf{\textit{TSC ResNet11}} & \textbf{\textit{CRNN}} \\
\midrule
random		& 78.5 (9.2) &  81.2 (9.2) &  10.1 (2.7) &  9.2 (2.6)  \\
pwgen		& 76.8 (7.9) &  76.9 (9.3) &  10.0 (2.0) &  9.0 (2.2)  \\
xkcd		& 73.0 (6.5) &  74.2 (6.2) &  10.8 (1.3) &  10.7 (1.6) \\
insecure	& 69.9 (10.0) & 72.0 (10.7) & 11.1 (2.8) &  11.3 (3.1) \\
\bottomrule
\end{tabular}
\end{table}

\subsubsection{\ac{EMG} vs \ac{IMU}}
\begin{figure}
\centering
\subfigure{\includegraphics{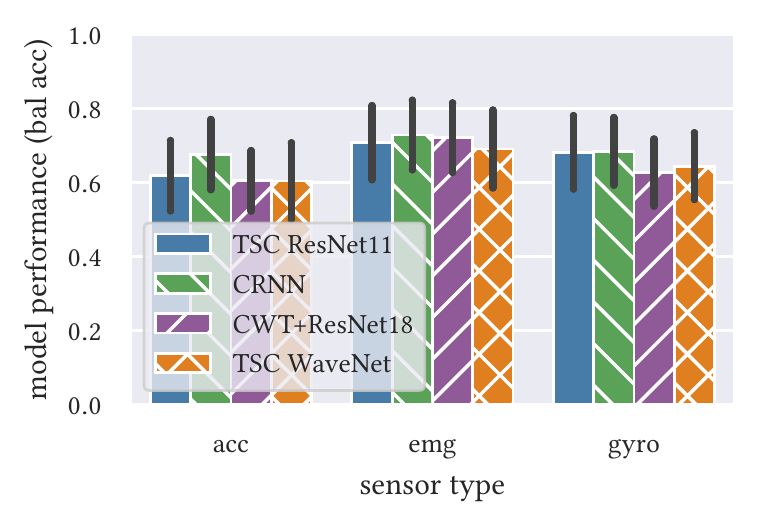}}
\Description{A bar plot showing the mean balanced accuracy of the different classifiers if only the accelerometer data is used (about \unit{60-66}{\%}), only the gyroscope data is used (about \unit{62-68}{\%}) or only the \ac{EMG} data is used (about \unit{70}{\%}). For the EMG data the results vary least between classifiers and for the accelerometer data the result is significantly better for the CRNN compared to the others. The standard deviation of all models is high.}
\subfigure{\includegraphics{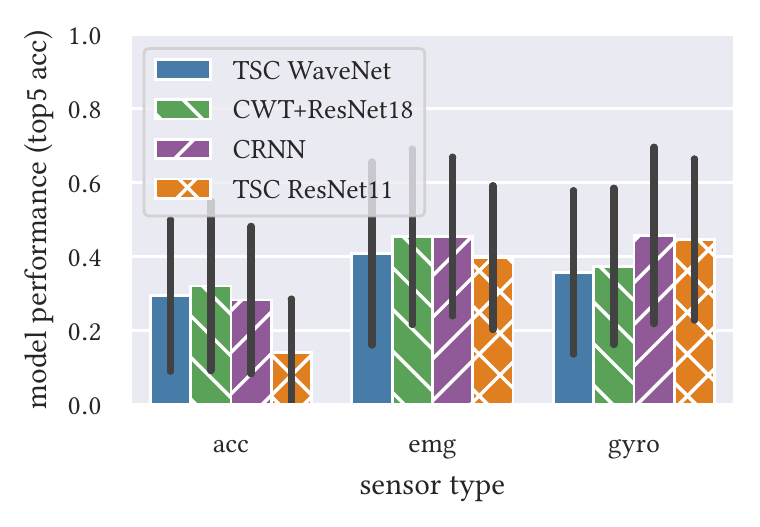}}
\Description{A bar plot showing the mean top-5 key accuracy of the different classifiers if only the accelerometer data is used (about \unit{12-30}{\%}), only the gyroscope data is used (about \unit{34-43}{\%}) or only the \ac{EMG} data is used (about \unit{40-42}{\%}). For the EMG data the results vary least between classifiers and for the accelerometer data the result is significantly worse for the TSC ResNet11 compared to the others. The standard deviation of all models is high.}
\caption{The mean balanced accuracy of the keystroke detection (left) and the mean top-5 key accuracy of the key identification (right) and the respective standard deviation for different sensor types and classifiers.}
\label{fig:sensor_comparison}
\end{figure}

To compare the performance of each model on the \ac{EMG}, accelerometer and gyroscope sensor, we train one model each using only the data of the respective sensor.
When training the models only on the \ac{EMG} data, the performance drops slightly (\unit{3.2}{\%} in the mean balanced accuracy for the CRNN) compared to training on all data. When training the models only on the gyroscope data, this performance drop increases to \unit{7.7}{\%} in the mean balanced accuracy for the CRNN and the results are less consistent among the four classifiers, with the TSC ResNet11 and the CRNN performing better than the other two, as shown on the left in Fig.~\ref{fig:sensor_comparison}. This effect is even more prominent for the accelerometer data for which the CRNN performs best with a performance drop of \unit{8.5}{\%} in the mean balanced accuracy compared to using all data, with the other classifiers showing drops as high as \unit{13.5}{\%} and only reaching performances of about \unit{60}{\%} mean balanced accuracy. This comparison suggests that the accelerometer data has little influence on the results and that the results mostly depend on the \ac{EMG} data.

\subsubsection{Known vs Unknown Participants/Passwords}
\begin{figure}
\centering
\subfigure{\includegraphics{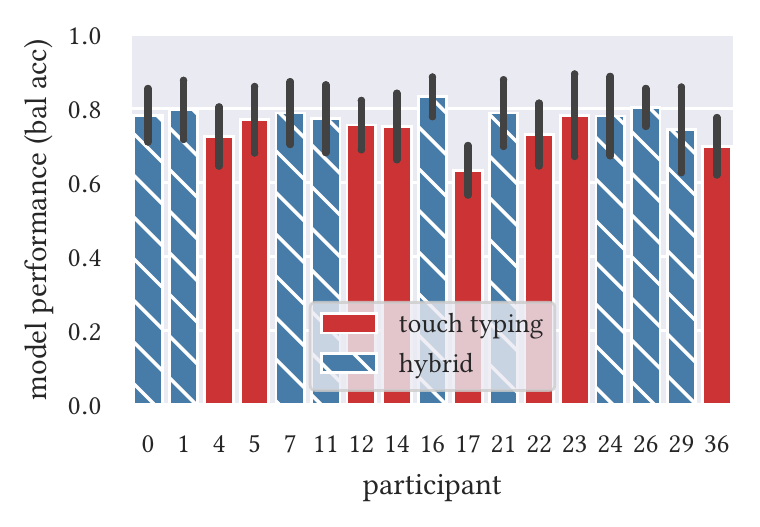}}
\Description{A bar plot showing the mean balanced accuracy of the CRNN on each participant, highlighting touch typists and hybrid typists. Values range from \unit{63}{\%} to \unit{83}{\%} mean balanced accuracy with most around \unit{78}{\%}. The hybrid typists show slightly better results than the touch typists.}
\subfigure{\includegraphics{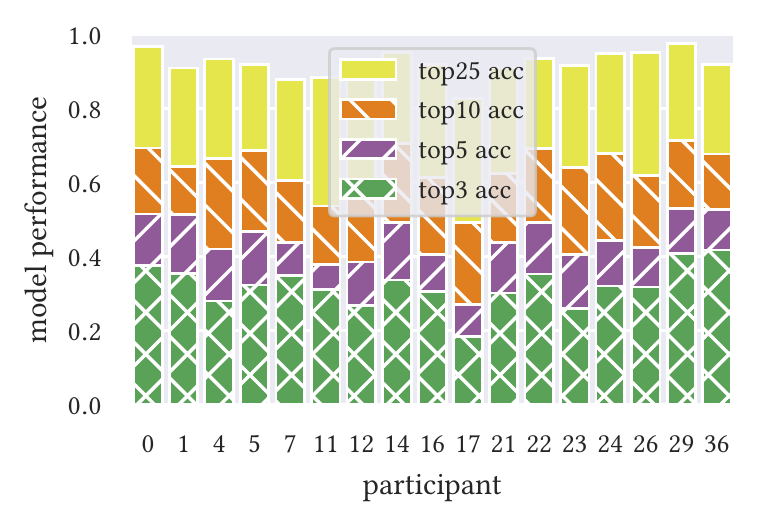}}
\Description{A stacked bar plot showing the mean top-3, top-5, top-10 and top-25 key accuracies of the \ac{CWT}+ResNet18 on each participant. The top-n accuracies vary between participants for different values of n. Values for the mean top-25 key accuracy range from \unit{50}{\%} to about \unit{98}{\%}. The mean top-10 key accuracy ranges from \unit{30}{\%} to about \unit{70}{\%}, the mean top-5 key accuracy ranges from \unit{20}{\%} to about \unit{55}{\%} and the mean top-3 key accuracy ranges from \unit{0}{\%} to almost \unit{50}{\%}.}
\caption{The mean balanced accuracy of the keystroke detection of all participants of the CRNN model (left) and the mean top-n key accuracies of the key identification of all participants of the \ac{CWT}+ResNet18 model (right). The standard deviation is omitted for better visibility on the right.}
\label{fig:participant_difference_detection}
\end{figure}

Comparing the average performances on \textit{known} participants whose data is present in the training data with the average performances on \textit{unknown} participants yields no significant difference compared with the standard deviation of both results. This may be due to the effect described in Section~\ref{sec:dataset}, concerning recordings of the same participant in a new recording session, but also shows the capability of the models to generalize between participants.
The same observation applies when comparing the results on shared data, i.e.\ passwords typed by multiple participants, with the results on individual data, i.e.\ passwords typed by only one participant each.
However, the generalization does not work equally well for all participants as the mean balanced accuracy for individual participants ranges from \unit{63.3}{\%} to \unit{83.2}{\%}, as shown on the left in Fig.~\ref{fig:participant_difference_detection}. Nevertheless, the results on each of the participants is significantly better than guessing, with the performances for many of them showing similar results. These differences are similar across all considered models.

\subsection{Key Identification Results}\label{sec:results_id}

Table~\ref{tab:results_identification} shows the average performance of each classifier on the entire test data, i.e.\ evaluated on all participants and all password types.
Note that, if not stated otherwise, the key identification is evaluated separately from the keystroke detection on samples containing a valid keystroke.
We thus give an estimation of the performance independent of our keystroke detection and possible improvements thereof, e.g.\ by incorporating further modalities or choosing different temporal tolerances for the post-processing step.

The \ac{CWT}+ResNet18 shows the best performance of about \unit{32.1}{\%} for the mean top-3 key accuracy, followed by the \ac{TSC} WaveNet with a mean top-3 key accuracy of \unit{27.8}{\%}.
The \ac{TSC} ResNet11 and CRNN model only reach a mean top-3 key accuracy of about \unit{24}{\%}.
All four models achieve a mean performance value significantly higher than random guessing ($3/52 \approx \unit{6}{\%}$ for the mean top-3 key accuracy, considering 52 classes) and for all models, the majority of measured performance values is higher than random guessing.
However, all of the performance values exhibit comparatively large variations, especially when compared to the keystroke detection.

\begin{table}
\caption{Mean and Standard Deviation of Top-n Key Accuracy Key Identification Results}
\label{tab:results_identification}
\begin{tabular}{ccccc}
\toprule
\textbf{Classifier Type} & \textbf{Top-3 Acc. (\%)} & \textbf{Top-5 Acc. (\%)} & \textbf{Top-10 Acc. (\%)} & \textbf{Top-25 Acc. (\%)}\\
\midrule
TSC ResNet11	& 24.0 (15.8) & 33.9 (18.7) & 53.4 (19.9) & 84.8 (14.0) \\
CWT+ResNet18	& 32.1 (20.3) & 44.4 (23.4) & 64.1 (25.3) & 92.2 (11.8) \\
CRNN			& 24.3 (17.6) & 33.8 (21.4) & 52.5 (23.5) & 81.7 (17.1) \\
TSC WaveNet		& 27.8 (20.4) & 39.4 (23.7) & 58.2 (26.0) & 87.4 (14.2) \\
\bottomrule
\end{tabular}
\end{table}

The right part of Fig.~\ref{fig:binary_temp_predictions} shows the impact of our binary classification with a peak detection for different temporal tolerances on the results of our multi-class classifier, compared to the improved keystroke detection rates when increasing the temporal tolerance.
We observe that the mean performance of our second stage is very similar for temporal tolerances of \unit{0-50}{\milli\second}, which indicates that our assumption made in Section~\ref{sec:postprocessing} is correct and that the multi-class performance results are invariant to a shift of a few samples between a true keypress event and its prediction.
Note that we only consider the true positives for this test of our second stage, as using the false positives would not allow a meaningful evaluation of the robustness against classifying keystrokes with slightly miss-aligned samples.

The right part of Fig.~\ref{fig:participant_difference_detection} shows the performance of the key identification on each participant, assuming a perfect keystroke detection.
For all of the participants, the mean top-n key accuracy is significantly above random guessing but the results vary more strongly between different participants than for the binary model on the left of Fig.~\ref{fig:participant_difference_detection}. Notably, both the binary and multi-class models perform worst on participant 17 for both the mean balanced accuracy as well as across all mean top-n key accuracies, but the best performance is not consistent between the binary and multi-class classifiers. Nevertheless, the results show that a generalization between users is possible, though the results will vary for some users.

\subsubsection{Password Types}
The right part of Fig.~\ref{fig:performance_speed_comparison} shows the performance of the multi-class \ac{CWT}+ResNet18 on different password types, an average of which is given in Table~\ref{tab:results_password_type_mul}. While the insecure passwords show a very high variation and a high number of samples for which the classifier has a mean top-3 key accuracy of \unit{0}{\%}, the results on xkcd passwords are comparatively good and show little variation, even though these are typed faster than the random and pwgen passwords which show a worse performance on average. This is contrary to the results for the binary classifier, which works best on slowly typed data, as shown on the left in Fig.~\ref{fig:performance_speed_comparison}. This may indicate that a consistency in the typing pattern, which is more likely for longer text than for random character combinations (and also for the entry of a well-known password), has more influence on the key identification results than the typing speed.

\begin{table}
\caption{Mean and Standard Deviation of Key Identification Results per Password Type (CWT+ResNet18)}
\label{tab:results_password_type_mul}
\begin{tabular}{ccccc}
\toprule
\textbf{Password Type} & \textbf{Top-3 Acc.} & \textbf{Top-5 Acc.} & \textbf{Top-10 Acc.} & \textbf{Top-25 Acc.} \\
\midrule
random		& 17.7 (12.8) & 27.9 (14.5) & 47.3 (17.9) & 84.7 (13.6) \\
pwgen		& 28.7 (15.0) & 41.3 (16.6) & 62.3 (14.0) & 90.9 (9.2)  \\
xkcd		& 47.5 (10.4) & 64.0 (10.5) & 84.4 (8.2) &  98.6 (2.0)  \\
insecure	& 34.6 (26.4) & 44.3 (30.6) & 62.4 (36.0) & 94.7 (13.6) \\
\bottomrule
\end{tabular}
\end{table}

\subsubsection{\ac{EMG} vs \ac{IMU}}
To compare the performance of each model on the \ac{EMG}, accelerometer and gyroscope sensor, we train one model each using only the data of the respective sensor. The models trained only on the \ac{EMG} data perform slightly better overall, being more consistent for different models than the ones trained only on gyroscope data, as shown on the right in Fig.~\ref{fig:sensor_comparison}. However, the best results based on the gyroscope by the \ac{TSC} ResNet11 and the CRNN are similar to the best result on the \ac{EMG} data by the \ac{CWT}+ResNet18 and the CRNN. Both results are significantly better than the results for the accelerometer, for which the \ac{TSC} ResNet11 comes close to guessing. All results show a comparatively high variation, in particular compared to the binary results on the left of Fig.~\ref{fig:sensor_comparison}. These results show that the \ac{EMG} and gyroscope sensor data are about equally important for the key identification. However, depending on the classifier used, the accelerometer negatively influences the performance when combining all sensors, with the \ac{EMG} and gyroscope data individually yielding a better result for the \ac{TSC} ResNet11 and the CRNN.

\subsection{Result Overview}
In summary, we apply four different end-to-end machine learning models for (1) keystroke detection and for (2) key identification.
The \ac{CRNN} and the \ac{TSC} ResNet11 perform best for keystroke detection with a mean balanced accuracy of \unit{76.1}{\%} and \unit{74.6}{\%} respectively.
Applying a peak-detector in combination with a temporal tolerance allows us to boost this performance even further depending on the chosen tolerance.
For example, a conservative choice of \unit{25}{\milli\second}, which is smaller than \unit{99}{\%} of all keystroke intervals observed in our training data, boosts the mean balanced accuracy to \unit{86.1}{\%} and the f1 score to \unit{65.8}{\%}.
Depending on the speed of the typist, one could safely choose even larger values, which would synergize with our insights from Fig.~\ref{fig:performance_speed_comparison}, showing a higher vulnerability of slower typist.
As shown on the right in Fig.~\ref{fig:binary_temp_predictions}, the choice of the temporal tolerance does not influence the key identification.

For key identification the \ac{CWT}+ResNet18 performs significantly better than the other networks.
Assuming a perfect keystroke detection, it achieves a mean top-3 key accuracy of about \unit{32.1}{\%} when classifying 52 different keys as part of our key identification.
In contrast to the keystroke detection, the key identification performance is mostly independent of the typing speed but may be influenced strongly by the typing pattern and its consistency. Passwords such as the xkcd password based on multiple, randomly selected words to achieve a high entropy seem to be among the most vulnerable for our approach.

In comparison, our \ac{TSC} WaveNet model is consistently outperformed in both keystroke detection and key identification, though for the latter it performs better than the \ac{CRNN} and \ac{TSC} ResNet11 on all data. In general, all four models proved to be able to learn patterns from the raw data and yield results better than guessing.
Taking a look at real-world password entry, a well-known password will usually by typed fast, which is disadvantageous for our keystroke detection, but it will presumably also always be typed in a similar way due to muscle memory, for which in turn our key identification performs well, as shown for the xkcd passwords on the right in Fig.~\ref{fig:performance_speed_comparison}.

For the keystroke detection, we observe that the \ac{EMG} data is the most important sensor, closely followed by the gyroscope, both of which are better than the accelerometer. For the key identification the \ac{EMG} data is similarly important to the gyroscope data and more so regarding cross-model consistency of results, whereas the accelerometer data can negatively influence the performance depending on the choice of classifier. This supports our intuition that the \ac{EMG} data provides a more fine-grained insight into the movements of the finger compared to the \ac{IMU} data of the Myo armband and suggests that the \ac{EMG} and gyroscope sensors should be preferred when investigating the impact of a keylogging side-channel attack on a wearable worn on the lower arm.

Furthermore, we observe that for both the keystroke detection and the key identification, a generalization between users is possible.

%%%%%%%%%%%%%%%%%%%%%%%%%%%%%%%%%%%%%%%%%%%%%%%%%%%%%%%%%%%%%%%%%%%%%%%%%%%%%%%%%%%%%%%%%%%%%%%%%%%%

\section{Limitations and Discussion}\label{sec:discussionlimits}

Though we tried to replicate a realistic scenario, there are some limitations to our data and approach.

First, we have used only one keyboard for all recordings, thus we cannot provide sufficient data about the generalization capabilities of our approach concerning different models of keyboards.

Second, the participants wore the Myo armband always in a specific position and orientation on the lower arm. Since this position is predefined by the vendor, users of the Myo armband will most likely use the same positioning. But creating a more extensive dataset with multiple Myo armband positions may result in a more powerful model, in particular when regarding possible generalizations across devices such as prostheses. Including a wrist-worn wearable to directly compare our findings regarding the performance influence of the different sensors would also be interesting in the future.

Third, as mentioned in Section~\ref{sec:datastudy}, wearables constantly generate data which is not directly linked to the action of typing on a keyboard. Thus, an additional typing activity detection should be explored to reduce the number of false predictions of keystrokes during different daily activities. As this also provides a possible mitigation for the presented attack by cutting out the detected sequences of typing from the sensor data, this would be a worthwhile addition for future work.

Forth, while we include the \texttt{shift} modifiers in our dataset, our classification approach is unable to detect if a pressed \texttt{shift} modifier applies to one or multiple keys, as we do not consider key releases. An example of this is shown in Fig.~\ref{fig:binary_prediction_example} where \texttt{shift} is pressed to type both \textit{/} and \textit{R} on a German keyboard layout. A multi-label approach with a key state encoding or a combination of classifiers predicting press and release events could solve this.

Fifth, our classifiers lack knowledge about predictions on neighboring samples and thus are unable to learn that keypresses usually do not appear in consecutive samples. Including the prior truth into the models or choosing a sequence-to-sequence approach may improve the results and make the peak detection obsolete.

Apart from the restrictions described above, we have tried to simulate natural typing behavior by including \texttt{backspace} and the \texttt{shift} modifier, including different postures occurring in everyday keyboard usage, as well as allowing non-typing movements, small breaks and individual typing styles. We have included an extensive set of keys of a physical keyboard, focusing on the physical keys rather than their character representation in order to be independent of a logical keyboard layout. Furthermore, we present a dataset of diverse training and test data meant to mimic realistic password entry in order to evaluate keylogging side-channel attacks based on \ac{EMG} and \ac{IMU} data recorded from the lower arm.

Our results show that even under these difficult conditions, significant information can be gained regarding the inference of keystrokes from sensor data. According to our experiments, the \ac{EMG} data in particular can raise the performance of such inferences and could be used without requiring additional sensor data or as a surrogate for the accelerometer or for microphones in approaches that combine sensor data for key identification with audio data for keystroke detection.
Compared to previous attacks focusing on \ac{IMU} devices, typically worn on the wrist, \ac{EMG} devices offer an additional attack vector, but also a potentially better way to infer keystrokes as indicated by our experiments.

Furthermore, it may be interesting to further explore the potential of using neural networks based on end-to-end learning for between-subject approaches. Creating more open access datasets would be a next step towards enhancing and comparing such approaches. If each new sensor modality opens up new side channels, this possibility should be explored in realistic and replicable settings to be able to explore feasible mitigations.

\section{Related Work}\label{sec:relatedwork}

Multiple papers have investigated wrist-worn smartwatches or fitness trackers to infer (pass)words typed on a physical keyboard \cite{wang2015mole, liu2015good, maiti2016smartwatch, pandelea2019password} or \acp{PIN} typed on a number pad like a payment terminal used in stores \cite{liu2015good, wang2016friend, liu2018aleak}.
All this work relies on motion data captured solely by an accelerometer and a gyroscope \cite{wang2015mole, liu2018aleak, pandelea2019password}, an accelerometer and a magnetometer \cite{wang2016friend}, or an accelerometer while using the microphone to detect keystrokes \cite{liu2015good, maiti2016smartwatch}.

Similar research has been done to infer passwords typed on the touchscreen-based virtual keyboards of smartphones \cite{sarkisyan2015wristsnoop, maiti2015smart, maiti2018side}, but the motion when typing on a smartphone is very different from that on a physical keyboard and often dominated by the thumb.
The latter is used mostly for typing the \texttt{space} key on a physical keyboard and is often disregarded by some of the work related closely with ours because the authors only consider words or passwords (e.g.\ \cite{wang2015mole, Zhang2017}).

Of those targeting a physical keyboard, most include a language analysis and focus on recovering words or sentences from the English language \cite{wang2015mole, liu2015good, maiti2016smartwatch}. Only \citet{pandelea2019password} mention passwords, including alphabetic characters and digits. But they do not reach a competitive accuracy for the full keyboard, only on the number pad. Most enforce or assume the user to use touch typing \cite{wang2015mole, liu2015good, pandelea2019password}, only \citet{maiti2016smartwatch} do not.

\citet{pandelea2019password} are the first to apply a convolutional neural network to the problem of inferring keystrokes. But the authors use it on features and have not tried end-to-end learning with neural networks.

\citet{yang2018armin} implement a partial text-entry system using five channels of \ac{EMG} data acquired from the left forearm with a custom system for predicting 16 different keys with feature-based classifiers.
The authors use a custom algorithm to determine the onset and offset of keystrokes by comparing the output value to a user-dependent threshold value.
Similar to the basic idea of our peak detector, the authors revise their prediction by taking the typical timing of a keystroke into account.
For doing so, they discard recognized keystrokes with a duration smaller than \unit{300}{\milli\second}.
However, this is a larger value than encountered for the majority of keystrokes in our training data as shown on the right in Fig.~\ref{fig:pp_intervals}.

\citet{Zhang2017} are closest to our work as they also use the Myo armband and are the only ones apart from us using \ac{EMG} data for a keylogging side-channel attack on a physical keyboard. The authors describe both an attack to infer \acp{PIN} from typing on a number pad as well as inferring passwords from typing on a physical keyboard.
They use a custom algorithm for detecting keystrokes using the \ac{EMG} data collected from the Myo armband.
For key identification, the authors use a feature-based finger classification based on the \ac{EMG} data in combination with a hand position tracking derived from the accelerometer data to identify keystrokes in passwords consisting of four to eight characters recorded as test data.

The authors assume that the target of the described attack uses the Ratatype typing style, a variant of touch typing, and that the data is recorded in a strict environment, i.e.\ the participants in their experiments were told to avoid movements and to keep their wrists above the table while typing \cite{Zhang2017}. Both are strong assumptions which limit the generalizability of the approach, as many typists frequently change their posture and move their hands or arms away from the keyboard or table. One could argue that this is more likely to happen given a natural break, i.e.\ at the end of a sentence, and less likely to happen while the users type a word, in particular a password. But taking a different posture may have an impact on the classification itself, as does the typing style, and given that the authors assume perfect Ratatype typing, their classifier will not be able to classify other typing styles correctly. However, many other typing styles exist and even most touch typists do not display a perfect finger-to-key mapping \cite{feit2016we}.

In our work, we have tried to minimize assumptions and assume a realistic scenario in order to evaluate the feasibility of an actual attack on a user. The results show that our approach is barely influenced by a person's typing style with our classifiers performing slightly better on hybrid typists. We have gathered an extensive dataset and tested our classifiers on different types of passwords of varying lengths and strengths. Furthermore, we have compared our approach on both the \ac{EMG} and \ac{IMU} data and found that it works best for the \ac{EMG} and worst for the accelerometer data, which also makes it a valid option to only rely on the \ac{EMG} data.

Furthermore, \citet{Zhang2017} use a within-subject approach, which requires recording of and training a classifier on a victim's data to launch the attack.
This approach requires either to extract both the sensor data and the ground truth (supervised learning) or to extract only the sensor data and to apply an unsupervised learning model.
In case of the key identification, this could be achieved by using a language model to leverage known structures for predicting the most likely keys.
However, it does not allow generalizing between users, i.e.\ training a classifier on a set of users before targeting victims outside this set.
In fact, the authors claim that they have tried a generalizing approach but have not succeeded due to the differences in different user's muscle structures.

We have analyzed the data in our dataset and found that even under the assumption that participants wear the Myo armband in the same way on different days, the recordings of different sessions differ significantly, being in part more similar to those of other participants, as shown in Fig.~\ref{fig:within_between_user_quetzalcoatl}.
This may render a within-user approach useless if the classifier has to be retrained each time the Myo armband is taken off.
We have further shown that a between-subject approach works well on most users, making it a powerful and feasible option for further exploration of using \ac{EMG} data in keylogging side-channel attacks.

In general, it is hard to compare different approaches, as, for example, smartwatches are usually worn on the wrist while the Myo armband is worn on the lower arm.
Furthermore, metrics such as accuracy and recall convey little meaning by themselves in a keystroke detection scenario which is heavily influenced by class skew, as shown in Section~\ref{sec:dataset} and as illustrated for the accuracy on the left in Fig.~\ref{fig:binary_temp_predictions}. Similarly, metrics such as key accuracy, word accuracy or metrics that rely on the look-up of a word in a given dictionary cannot be directly translated to each other.
In order to improve comparability, multiple metrics need to be reported and open access datasets have to be created, against which different approaches can be measured.
With this work, we have attempted this while investigating a new sensor modality and bridging the gap to existing keylogging side-channel attacks by including and comparing the \ac{IMU} data of lower arm movement in our dataset and evaluation.

%%%%%%%%%%%%%%%%%%%%%%%%%%%%%%%%%%%%%%%%%%%%%%%%%%%%%%%%%%%%%%%%%%%%%%%%%%%%%%%%%%%%%%%%%%%%%%%%%%%%

\section{Conclusion}\label{sec:conclusion}

In this work, we have explored the potential of using \ac{EMG} and \ac{IMU} data for a keylogging side-channel attack. We have implemented the first between-subject feature-less approach based on neural networks and \ac{EMG} data and evaluated it on a dataset of sensor and key data recorded in a realistic scenario using the Myo armband. We have shown the feasibility of such an approach and the advantage that can be gained by using the \ac{EMG} data to improve the performance of keylogging side-channel attacks, in particular compared with the accelerometer data.
Furthermore, we have created an extensive dataset which is available as open access along with the source code of this research.

\begin{acks}
We thank our shepherd and our anonymous reviewers for their insightful comments that helped to improve this paper.
Calculations for this research were conducted partially on the Lichtenberg high performance computer of the TU Darmstadt.
This work has been co-funded by the German Research Foundation as part of project C.1 within the RTG 2050 ``Privacy and Trust for Mobile Users'' as well as by the German Federal Ministry of Education and Research and the Hessian Ministry of Higher Education, Research, Science and the Arts within their joint support of the National Research Center for Applied Cybersecurity ATHENE.
\end{acks}

\bibliographystyle{ACM-Reference-Format}
\bibliography{bibliography}

\appendix

\section{Dataset and source code}

The dataset created as part of this work is available as open access at \url{https://doi.org/10.5281/zenodo.5594651}. The source code to record the data, train the models and reproduce our final results is available as free software at \url{https://github.com/seemoo-lab/myo-keylogging}.

\section{Classifier configuration details}\label{sec:appendixclassifiers}

The following describes the neural networks in more detail.

\subsection{TSC ResNet11}

% architecture in 2-3 sentences
Compared to the original architecture, due to using grouped convolutions, we use $\{8, 16, 16\}$ filters per input channel, totaling in $\{224, 448, 448\}$ filters based on our hyperparameter optimization, compared to $\{64, 128, 128\}$ in the original work.

\subsection{CWT ResNet18}

Our main changes to the original ResNet18 architecture are the removal of the max pooling layer and the reduction of the initial number of convolution filters from 64 to 32.

\subsection{CRNN}

% architecture
We use grouped convolutions to apply 16 individual filters per sensor channels (a total of 448 filters) with a kernel size of two.
Both \ac{LSTM} layers consist of 64 units and we apply a dropout of 0.4 for regularization.

\subsection{TSC WaveNet}

% architecture
The core architecture can be divided into three parts: The first part consist of a batch normalization layer followed by a convolution with kernel size one for adapting the number of input channels to the second part.
The second part of the model consists of a series of residual blocks as described in the work of \citet{Oord2016}.
In contrast to the models based on the ResNet architecture, the output of each residual block, which contain the dilated convolutions, is not only fed into the next block, but added together before being fed into the third part of the network.
The third part consists of a simple series of a \ac{ReLU}, a convolution with kernel size one and another \ac{ReLU}, followed by a convolution with kernel size one, which is connected to the output layer.

\end{document}